\documentclass[prb,aps,twocolumn,amssymb,showpacs]{revtex4}

\usepackage{epsfig}
\usepackage{graphicx}
\usepackage{dcolumn}
\usepackage{bm}

\begin{document}

\title{Circulating-current states and ring-exchange interactions in cuprates}

\author{     B. Normand }

\affiliation{D\'epartement de Physique, Universit\'e de Fribourg, 
             CH-1700 Fribourg, Switzerland}

\author{     Andrzej M. Ole\'s }

\affiliation{Marian Smoluchowski Institute of Physics, Jagellonian 
             University, Reymonta 4, PL-30059 Krak\'ow, Poland}

\date{\today}

\begin{abstract}

We consider the consequences for circulating-current states of a cyclic, 
four-spin, ``ring-exchange'' interaction of the type shown recently to 
be significant in cuprate systems. The real-space Hartree-Fock approach 
is used to establish the existence of charge-current and spin-current 
phases in a generalized Hubbard model for the CuO$_2$ planes in cuprates. 
We augment the Hartree-Fock approximation to consider correlated states 
renormalized by Gutzwiller projection factors, which allow us to gauge 
the qualitative effects of projection to no double occupancy in the 
limit of strong on-site Coulomb repulsion. We find that charge flux 
states may be competitive in the underdoped and optimally doped regime 
in cuprate systems, whereas spin flux states are suppressed in the 
strongly correlated regime. We then include the ring-exchange interaction 
and demonstrate its effect on current-carrying states both at and away 
from half-filling.

\end{abstract}


\maketitle

\section{Introduction}

The possibility of phases supporting circulating currents in two-dimensional 
(2D) systems has been debated for some time. The orbital antiferromagnet, or 
charge flux (CF) phase, was first identified in the context of excitonic 
semiconductors\cite{rhr} and reintroduced very early in the analysis of 
high-temperature superconductors,\cite{ram,rk} while the spin flux (SF) 
phase was introduced within a general planar model for the 
latter.\cite{rnvjk} On a square lattice, the CF phase [Fig.~1(a)], 
has counter-rotating currents of carriers on neighboring plaquettes, whence 
the alternative terminology ``staggered flux'' phase, and preserves SU(2) 
spin symmetry but breaks time-reversal invariance. The SF phase [Fig.~1(b)],
which can be said to be a fermionic realization of a spin nematic, may be 
considered as superimposed counter-rotating current patterns in which up- 
and down-spin currents are equal and opposite (no net charge transport), 
and preserves time-reversal invariance but violates SU(2) spin symmetry.

A considerable body of theoretical work now exists concerning these states, 
particularly the CF phase. Early analyses employed the Heisenberg-Hubbard
model with an SU($N$) generalization of the spin in order to obtain a 
controled mean-field decoupling,\cite{ram,rk} while further studies 
have since been performed of the different saddle points obtainable in 
the $t$-$J$ model.\cite{rvzs} The relationship of the CF phase to the 
superconducting one has been explored in detail within an SU(2)-symmetric 
slave-boson gauge theory of the the $t$-$J$ model by Lee and 
coworkers,\cite{rlnnw} and by using a Hubbard $X$-operator 
formulation.\cite{rcz} More general models for the cuprate planes, 
including not only a finite on-site Hubbard interaction ($U$), but also 
explicit nearest-neighbor superexchange ($J$) and Coulomb interaction ($V$) 
terms, have been considered using weak-coupling techniques. At half-filling
and for $U > 0$, the Hartree-Fock approximation\cite{rnvjk} (HFA) and 
more sophisticated renormalization-group analyses of divergent 
susceptibilities\cite{rbbd,rkk1} agree on a phase diagram with magnetic 
order for weak $J$ and $V$, a charge-density-wave (CDW) phase for strong 
$V > 0$, CF for $V > 0$ and sufficiently large $J > 0$, and SF for 
$V > 0$ and $J < 0$. Negative $V$ (nearest-neighbor attraction) yields 
superconducting phases of singlet symmetry if $J > 0$ and triplet symmetry 
if $J < 0$.\cite{rmrr}

\begin{figure}[t!]
\mbox{\hspace{0.3cm}\includegraphics[width=3.4cm]{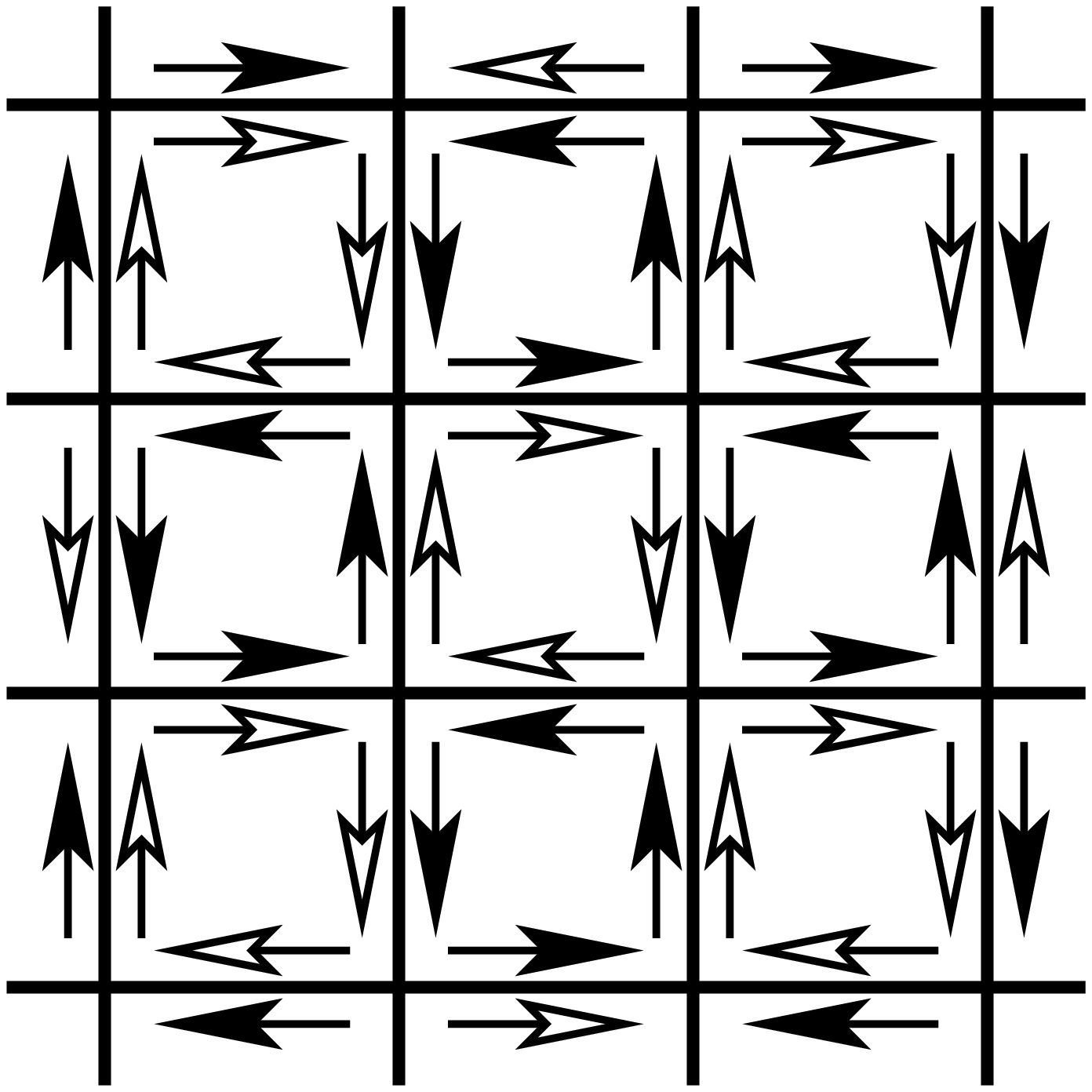}}
\mbox{\hspace{0.3cm}\includegraphics[width=3.4cm]{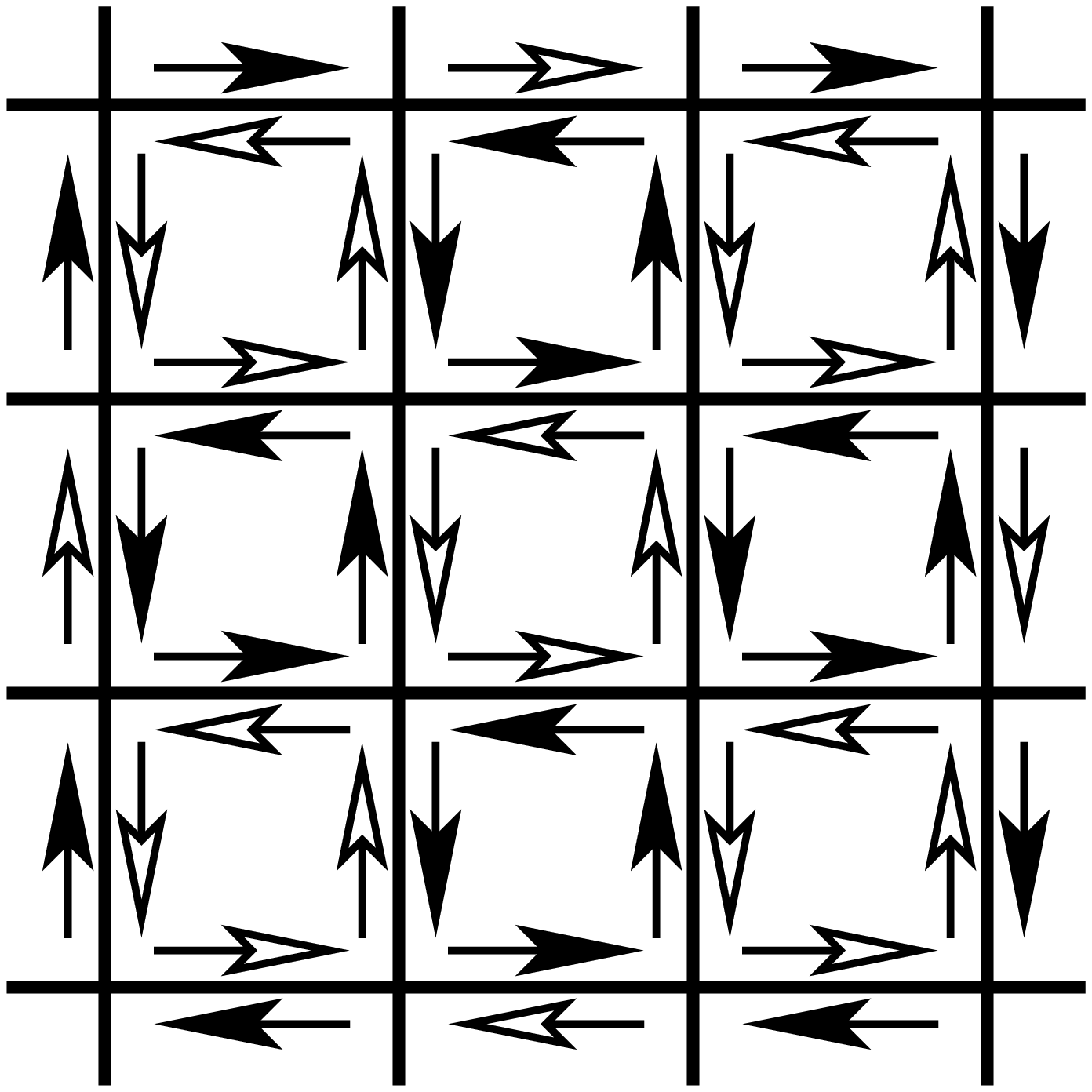}}
\centerline{ \quad (a) \qquad\qquad\qquad\qquad\qquad (b) }
\caption{ Schematic representation of (a) charge-flux (CF), and (b) 
spin-flux (SF) phases. Solid arrows represent currents of spin-up 
electrons and open arrows spin-down electrons. }
\end{figure}

Interest in the CF phase has been revived recently by an accumulation 
of experimental evidence suggesting the presence of a quantum critical 
point in the superconducting region of the phase diagram.\cite{rtlp} 
One of the proposals for the ``hidden'' order parameter associated 
with the transition is the conventional CF phase, also known as the 
$d$--density--wave (DDW) state,\cite{rclmn} although more complex current 
patterns based on a three-band description have also been proposed.
\cite{rvsv} To date the only experimental means of identifying a CF state 
is the breaking of time-reversal symmetry, and the interpretation of one 
such measurement using polarized angle-resolved photoemission 
spectroscopy\cite{rkrfclrcyovh} remains open. Circulating-current 
phases have been considered numerically in the context of exact 
diagonalization (ED) studies of chiral fluctuations in the 2D $t$-$J$ 
model,\cite{rpdr} and fluctuating staggered currents on a doped 
ladder,\cite{rtpc} while a more general model displaying steady 
currents was considered by density-matrix renormalization-group (DMRG) 
studies in a ladder geometry.\cite{rmfs}

The possible significance of cyclic four-spin, or ``ring-exchange'', 
interactions in cuprates was also recognized at an early stage.\cite{rcgb}
The important term arises at fourth order in a strong-coupling (small 
$t/U$) perturbative expansion of the single-band Hubbard model,\cite{rtmgy} 
and has been shown in this limit to give the leading correction to the 
nearest-neighbor Heisenberg model. Experimentally, inelastic neutron 
scattering measurements of the spin-wave spectrum in the 2D system 
La$_2$CuO$_4$,\cite{rcmapfmcf} and in the quasi-one-dimensional spin-ladder 
compound La$_6$Ca$_8$Cu$_{24}$O$_{41}$,\cite{rmkebm} as well as two-magnon 
Raman scattering measurements on doped systems related to the 
latter,\cite{rws} indicate the presence of ring-exchange contributions. 
The magnitude of this interaction has been quantified as being 20\% of the 
nearest-neighbor superexchange interaction, i.e.~$K \simeq J/5$, both by 
systematic expansion of the three-band Hubbard model,\cite{rmhr} and by 
a self-consistent spin-wave analysis yielding good quantitative agreement 
with the experimental data.\cite{rkk2}

At the qualitative level, the presence of a ring-exchange term has been 
found by consideration of exactly soluble extended Heisenberg models to 
generate a rich variety of novel phases on a 1D spin ladder.\cite{rgnb}
A number of these phases have been found and characterized by ED and DMRG 
studies of the isotropic ladder.\cite{rlst} However, for 2D systems 
accurate results beyond the Hartree-Fock analysis of static magnetic 
phases\cite{rcgb} remain scarce, and the only numerical study to date was 
performed for a system with XY spin interactions.\cite{rsdss}

The ring-exchange interaction on a square lattice may be considered as 
the product of a virtual hopping process around an elementary plaquette, 
while the CF and SF phases are those in which real currents circulate on 
the plaquettes. However, to our knowledge there have been rather few 
studies of the mutual effects of flux states and ring exchange. The 
issue has been addressed for the CF phase of a $t$-$J$ model with 
unprojected nearest- and next-neighbor hopping by applying a large-$N$ 
approach.\cite{rckk} This study suggested that a ring-exchange interaction 
acts to suppress the homogeneous CF phase, but may favor inhomogeneous 
current patterns and other states involving a further breaking of 
translational symmetry.\cite{rsdss}

Here we adopt a qualitative approach to gain further physical insight 
into both the real-space nature of circulating-current phases on the 2D 
square lattice and the influence upon them of ring-exchange interactions. 
In Sec.~II we present the generalized Hubbard Hamiltonian supporting 
flux phases in the weak-coupling limit and discuss its treatment in 
the HFA. Sec.~III contains a summary of the properties of the CF and SF 
phases obtained in HFA with unprojected hopping. We turn in Sec.~IV 
to a study of electronic correlation effects, which are included within 
the Gutzwiller approximation through projected hopping terms, and this 
allows us to make contact with the physical parameter regime for 
cuprate systems. In Sec.~V we include the ring-exchange interaction and 
demonstrate its effects on both CF and SF phases at different fillings 
by analyzing the plaquette current and the ring-exchange energy 
contribution. Sec.~VI contains a summary and conclusions.

\section{Model and Formalism}

\subsection {Extended Hubbard model}

To establish the existence of CF and SF phases (Fig.~1) in a 2D lattice 
we consider the extended Hubbard model in the form 
\begin{equation}
{\cal H} = H_{t} + H_{U} + H_{J} + H_{V} + H_{K}, 
\label{eehh} 
\end{equation}
where the first two terms
\begin{eqnarray}
H_{t} & = & 
- t\sum_{\langle i,j \rangle \sigma } 
  ( c_{i\sigma}^{\dag}c_{j\sigma} + {\rm H. c.})         
- t'\sum_{\langle\langle i,j \rangle\rangle \sigma } 
  ( c_{i\sigma}^{\dag}c_{j\sigma} + {\rm H. c.}),         \nonumber \\
H_{U} & = & U \sum_i n_{i \uparrow} n_{i \downarrow} , 
\label{ehh} 
\end{eqnarray}
constitute the conventional Hubbard Hamiltonian for a single, nondegenerate 
band with nearest- and next-neighbor hopping. Here $c_{i\sigma}^{\dag}$ 
denotes the electron creation operator, $n_{i\sigma} = c_{i\sigma}^{\dag} 
c_{i\sigma}$, $n_i = \sum_{\sigma} n_{i\sigma}$, and the notatation 
$\langle i,j \rangle$ ($\langle\langle i,j \rangle\rangle$) indicates a 
pair of (next-)nearest-neighbor sites. We adopt $t = 1$ as the unit of 
energy. The terms 
\begin{equation}
H_{J} + H_{V} = J \sum_{\langle i,j \rangle} {\bf S}_i {\bf \cdot S}_j 
+ V \sum_{\langle i,j\rangle,\sigma\sigma'} n_{i\sigma} n_{j\sigma'} ,
\label{ejv} 
\end{equation}
correspond, respectively, to superexchange and Coulomb interaction between
nearest neighbors. The $t$-$U$-$J$-$V$ model (\ref{eehh}) has been 
considered by a number of authors, including a Hartree-Fock treatment 
of the infinite system with only nearest-neighbor hopping\cite{rnvjk} 
and RG studies of both the half-filled case\cite{rbbd} with $t' = 0$ 
and the model with small $t'$ near the van Hove fillings.\cite{rkk1} 

Finally, the ring-exchange term in Eq.~(\ref{eehh}) takes the form
\begin{eqnarray}
\label{ehk}
H_{K} & \!\! = \!\! & K \sum_{\Box} \Big[ ({\bf S}_i 
{\bf \cdot S}_j) ({\bf S}_k {\bf \cdot S}_l) + ({\bf S}_i {\bf \cdot S}_l) 
({\bf S}_j {\bf \cdot S}_k)             \nonumber \\ & &
\;\;\;\;\;\;\;\; 
   -  ({\bf S}_i {\bf \cdot S}_k) ({\bf S}_j {\bf \cdot S}_l)\Big],
\end{eqnarray}
where $\Box$ denotes the sum over the four spins $(ijkl)$ of each plaquette
labeled clockwise. The phase diagram of a 2D system in the presence of this 
term has been considered only for the Heisenberg antiferromagnet,\cite{rcgb} 
although further progress has been possible for 1D magnetic 
systems.\cite{rgnb,rlst}

\subsection{Hartree-Fock Approximation}

The HFA is the simplest and the most efficient approach for the study 
of interacting electron systems. Although its application may seem 
inherently limited to the regime of weak interactions, it provides 
satisfactory results at all interaction strengths when the correlation 
energy is small (for example in magnetically polarized phases), and thus 
can be used to obtain meaningful qualitative insight into the nature of 
possible ordered phases also for large $U$. In the HFA, multiparticle 
interaction terms are decoupled into all possible operator pairs, all 
but one of which are replaced by $c$-number expectation values to leave 
a single-particle Hamiltonian. In the present analysis the choice of 
operator pairs with finite expectation values is as follows. Because our 
primary interest is the study of flux phases, we will work with a repulsive 
intersite Coulomb interaction $V > 0$, in which case no pairing decouplings 
need be considered, $\langle c_{i\sigma}^\dag c_{j\bar{\sigma}}^\dag \rangle 
= 0 = \langle c_{i\sigma}^\dag c_{j\sigma}^\dag \rangle$. In the absence 
of single-site spin-flip terms in the Hamiltonian [Eq.~(1)], there is no 
reason to expect finite averages of the form $\langle c_{i\sigma}^\dag 
c_{j\bar{\sigma}}\rangle$, where $\bar{\sigma}$ denotes the spin projection 
opposite to $\sigma$. Thus flux phases will compete or coexist with magnetic 
and possibly charge-ordered phases, and the finite expectation values to be 
considered are the site charge 
\begin{equation}
n_i = \langle c_{i  \uparrow}^{\dag} c_{i  \uparrow} 
            + c_{i\downarrow}^{\dag} c_{i\downarrow} \rangle, 
\label{esc}
\end{equation}
the site magnetic moment 
\begin{equation}
{\bf m}_i = \langle {\bf S}_i \rangle = {\textstyle \frac{1}{2}} \!\! 
\sum_{\alpha,\beta = \uparrow,\downarrow} \!\! \langle c_{i\alpha}^{\dag} 
{\vec\sigma}_{i\alpha\beta} c_{i\beta}\rangle,
\label{esm}
\end{equation}
and the bond order parameters 
\begin{equation}
s_{ij\sigma} = \langle c_{i\sigma}^{\dag} c_{j\sigma} \rangle.
\label{esb}
\end{equation}
In a circulating--current state, the latter parameters are complex, 
$s_{ij\sigma}\equiv |s_{ij\sigma}|e^{\chi_{ij\sigma}}$, with the real 
part corresponding to the bond kinetic energy and the imaginary part to 
the bond current. 

For later reference we present the Hartree-Fock decoupling expressions 
for all of the interaction terms in Eq.~(\ref{eehh}) in the form in which 
they will be used here. The Hubbard term is approximated by 
\begin{eqnarray}
\label{edu}
H_{U}\!\!\! & \simeq &\!\! U \sum_i \left[ \langle n_{i \uparrow} \rangle 
n_{i\downarrow} + n_{i\uparrow}\langle n_{i\downarrow}\rangle - \langle 
n_{i \uparrow} \rangle \langle n_{i \downarrow} \rangle \right. 
        \nonumber \\ 
	&- & \left.\!\!\!\! \langle c_{i\downarrow}^{\dag} 
c_{i\uparrow}\rangle c_{i\uparrow}^{\dag} c_{i\downarrow}\!-\!\langle 
c_{i\uparrow}^{\dag} c_{i\downarrow} \rangle c_{i\downarrow}^{\dag} 
c_{i\uparrow}\!+\!\langle c_{i\uparrow}^{\dag} c_{i\downarrow} \rangle 
\langle c_{i\downarrow}^{\dag} c_{i\uparrow}\rangle\right]\!, 
\end{eqnarray}
where the signs follow from fermionic statistics. We define the 
effective bond order parameter
\begin{equation}
g_{ij\sigma} \equiv s_{ij\sigma} + {\textstyle \frac{1}{2}} 
s_{ij\bar{\sigma}}, 
\label{gij}
\end{equation}
in terms of which a decomposition of the $J$, $V$, 
and $K$ interactions gives the following one-body terms in the Hamiltonian 
matrix. For the superexchange interaction
\begin{eqnarray}
H_{J} & \simeq & - {\textstyle \frac{1}{2}} J \sum_{\langle ij 
\rangle, \sigma} \left( g_{ji\bar{\sigma}} c_{i\sigma}^{\dag} c_{j\sigma}
 + g_{ij\bar{\sigma}} c_{j\sigma}^{\dag} c_{i\sigma} \right) \label{edj} 
\nonumber \\ & & + {\textstyle \frac{1}{2}} J \sum_{\langle ij 
\rangle,\sigma} \left( \lambda_\sigma S_i^z c_{j\sigma}^{\dag} c_{j\sigma} 
+ \lambda_\sigma S_j^z c_{i\sigma}^{\dag} c_{i\sigma} \right) 
\\ & & \!\!\!\!\!\!\!\!\!\!\!\! + {\textstyle \frac{1}{2}} J \sum_{\langle 
ij \rangle,\sigma} \left[ (S_i^x \! - \! i \lambda_\sigma S_i^y) 
c_{j\sigma}^{\dag} c_{j\bar{\sigma}} + (S_j^x \! - \! i \lambda_\sigma 
S_j^y) c_{i\sigma}^{\dag} c_{i\bar{\sigma}} \right], \nonumber
\end{eqnarray}
where $\lambda_{\uparrow} = 1$, $\lambda_{\downarrow} = - 1$, and 
$\{g_{ij\sigma}$, $g_{ji\sigma}\}$ are complex quantities. For the 
nearest-neighbor Coulomb interaction
\begin{eqnarray}
H_{V} & \simeq & V \! \sum_{\langle ij \rangle, \sigma} \left( 
n_i c_{j\sigma}^{\dag} c_{j\sigma} + n_j c_{i\sigma}^{\dag} c_{i\sigma}
\right) \label{edv} \nonumber \\ & - & V \! \sum_{\langle ij \rangle, 
\sigma} \left( s_{ji\sigma} c_{i\sigma}^{\dag} c_{j\sigma} + 
s_{ij\sigma} c_{j\sigma}^{\dag} c_{i\sigma} \right).
\end{eqnarray}

Finally, the ring-exchange interaction is decoupled into single-particle 
terms multiplied by all possible combinations of the averages obtained by
contracting the remaining six operators,  
\begin{eqnarray}
H_{K} \!\! & \simeq & \!\! - {\textstyle \frac{1}{2}} K \!\! 
\sum_{\langle ij \rangle, \sigma} \! \left[ g_{ji\bar{\sigma}} 
c_{i\sigma}^{\dag} c_{j\sigma} \! + \! g_{ij\bar{\sigma}} 
c_{j\sigma}^{\dag} c_{i\sigma} \right] \!\! \sum_{\langle 
kl \rangle(\langle ij \rangle)} \!\! \langle {\bf S}_k {\bf \cdot S}_l 
\rangle \label{edk} \nonumber \\ & & \!\! + {\textstyle \frac{1}{2}} K \!\! 
\sum_{i,j(i),\sigma} \! \left[ (S_j^x \! - \! i \lambda_\sigma S_j^y) 
c_{i\sigma}^{\dag} c_{i\bar{\sigma}} \! + \! \lambda_\sigma S_j^z 
c_{i\sigma}^{\dag} c_{i\sigma} \right] \nonumber \\ & & \times \!\! 
\sum_{\langle kl \rangle(\langle ij \rangle)} \!\! \langle {\bf S}_k 
{\bf \cdot S}_l \rangle, 
\end{eqnarray}
with the intersite spin correlation function, determined within HFA, given by 
\begin{eqnarray}
\langle {\bf S}_k {\bf \cdot S}_l \rangle & = & \langle {\bf S}_k \rangle 
\langle {\bf S}_l \rangle - {\textstyle \frac{1}{2}} (s_{kl\uparrow} 
s_{lk\downarrow} + s_{kl\downarrow} s_{lk\uparrow}) \label{edsisj} \nonumber 
\\ & & - {\textstyle \frac{1}{4}} (|s_{kl\uparrow}|^2 + |s_{kl\downarrow}|^2).
\end{eqnarray}
The structure of the decomposition is similar to that of the superexchange 
term, although in this case the net contributions are quartic in $\langle 
{\bf S}_i\rangle$ and $\langle c_{i\sigma}^{\dag} c_{j\sigma'}\rangle$, 
so may be expected to be small at the Hartree-Fock level. Note that in 
Eqs.~(\ref{edj}-\ref{edk}) we have omitted the constant terms bilinear 
in the expectation values [analogous to the final term in each line of 
Eq.~(\ref{edu})].

\subsection{ Self-consistent calculations in real space }

We have performed calculations on a finite cluster in real space to 
establish the microscopic conditions under which the CF and SF phases 
are manifest as circulating currents of real electrons (or holes), which 
arise from a mismatch in hopping expectation values between the two 
directions along each bond in a coherent pattern. These bond expectation 
values, the site charges, and the site magnetic moments are computed for 
any set of interaction parameters and iterated to self-consistency. The  
HFA in real space has been used extensively for qualitative modeling of 
cuprate systems, and in particular for the analysis of charge-inhomogeneous 
stripe and other density-wave phases, which it tends to favor. A full review 
of the technique is presented in Ref.~\onlinecite{rnk}, and we summarize 
here those aspects of its application which are important for the current 
analysis.

Most importantly, the HFA suppresses fluctuation terms and thus favors 
(homogeneous or inhomogeneous) ordered states which minimize static 
interactions. Magnetic and charge-density interactions, which are 
zeroth-order in particle-hopping fluctuations, are enhanced 
relative to kinetic terms, and it is for this reason that the method 
tends to find ground states with inhomogeneous charge distributions 
and/or magnetic order. By contrast, methods such as the Hubbard 
$X$-operator and other large-$N$ approaches have a tendency to favor 
kinetic terms, as a result of which they do not show stripe phases and 
often produce flux-phase ground states. Thus a real-space Hartree-Fock 
analysis is not an appropriate framework for an unbiased discussion of 
energetic contributions and true ground states; in specific terms, the 
flux phases we investigate are frequently local rather than global 
energy minima. Mindful of this fact, we will restrict our considerations 
to qualitative statements concerning the nature of circulating-current 
phases and the relative effect upon them both of doping and of the 
different interactions, particularly the ring-exchange term.

We note also that the method is especially sensitive to the choice of 
starting state, as well as to boundary conditions and in some cases 
to commensurability effects due to the electron filling. A careful 
specification of these extrinsic variables is required in order to make 
meaningful comparisons yielding intrinsic electronic properties.\cite{rnk}
We will consider a system of size $N$ = 12$\times$12 sites with periodic 
boundary conditions; qualitatively similar results were obtained with 
systems of size 8$\times$8 and 16$\times$16, and we have established that 
flux states may be found for any rectangular system with even-length sides. 
Open boundary conditions lead to a suppression of circulating currents 
and to inhomogeneous edge values, and are not considered further. The 
establishment of global, as opposed to local, energy minima may be pursued 
as in Ref.~\onlinecite{rnk}, but for comparative studies of flux states
is not of prime importance (preceding paragraph). Finally, by considering 
the evolution of the results with electron filling we will demonstrate
that commensuration effects are absent and thus that the system size 
is sufficient for robust qualitative conclusions.

\section{Circulating-Current States}

\subsection{Dependence on doping}

We begin by establishing homogeneous flux phases in the generalized 
Hubbard model on a cluster. As expected from the weak-coupling results 
mentioned in Sec.~I,\cite{rnvjk,rbbd,rkk1} we find that the nearest-neighbor 
interactions $J$ and $V$ are of crucial importance: circulating-current 
states appear close to half-filling only for simultaneously large values 
($\gtrsim 1$) of the interaction ratios $|J|/U$ and $4V/U$. The CF phase, 
with finite currents of up- and down-spin electrons moving in the same 
direction [Fig.~1(a)], occurs when $J > 0$ and SF states, with counterflowing 
up- and down-spin currents, appear for $J < 0$. The parameter sets for 
flux-phase formation are dictated by the strong propensity within the 
real-space Hartree-Fock approach towards charge localization, driven by 
large values of $U$, charge inhomogeneity driven by large $V$, and phase 
separation driven by large $J$. In this section we characterize the 
dependence of the homogeneous CF and SF phases on the doping and on 
the magnitude of the interactions. 

\begin{figure}[t!]
\centerline{\includegraphics[height=7.0cm,angle=270]{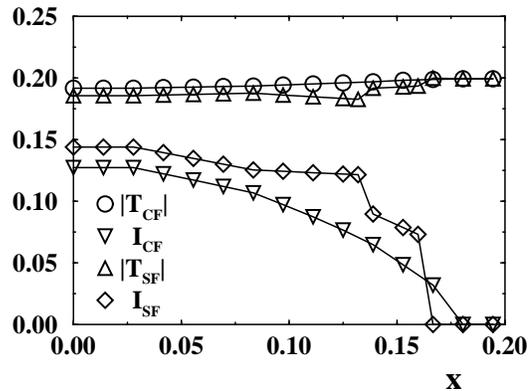}}
\caption{Doping-dependence of kinetic energy $T$ and current $I$ per bond 
for each spin orientation as obtained in HFA for CF ($J > 0$) and 
SF ($J < 0$) phases with $U = |J| = 4V = 4t$.}
\label{yicr}
\end{figure}

We define the filling as the normalized number of electrons in the system, 
$n = N_e/N$, and the hole doping as $x = 1 - n$. We choose the parameters 
$U = |J| = 4V = 4t$, and show in Fig.~\ref{yicr} the doping dependence of 
the bond order parameter [Eq.~(\ref{esb})] for a single spin species and 
a single direction along an $x$- or a $y$-axis bond. The real part 
$T = {\rm Re}\;s_{ij\sigma}$ is a component of the bond kinetic energy, 
and constitutes one eighth of the hopping kinetic energy per site of the 
system, while the imaginary part $I = {\rm Im}\;s_{ij\sigma}$ corresponds 
to one half of the bond current for a single spin species (and thus to one 
quarter of the charge current in a CF phase). It is clear that, for moderate 
interaction strengths, circulating-current states are robust close to 
half-filling. However, the current shows a monotonic decrease as a 
function of doping until the CF and SF phases become unstable at doping 
values of 0.16-0.18. 

The magnitudes of the (negative) real parts $T_{\rm CF}$ and $T_{\rm SF}$ 
are scarcely affected by the presence of the currents. In the absence of 
flux phases or other types of order, $T$ is expected to be constant at low 
doping for the cluster under consideration, beginning to drop beyond $x = 
0.153$; this number does not affect the behavior of established ordered 
phases. For the intermediate coupling strengths used in Fig.~\ref{yicr}, 
the kinetic contributions in the CF and SF phases are suppressed when the 
currents are maximal. This reduction is by approximately 5\% below the 
value found in the order-free ``metallic'' phase realized beyond the 
critical doping $x \sim 0.17$. We comment on the fact that the current 
``order parameter'', $I_{\rm CF}$, drops continuously to zero with 
increasing doping, while $I_{\rm SF}$ changes discontinuously. This 
contrasting behavior may be taken as a reflection of the different 
symmetries broken by the two phases (Sec.~I), through the coexistence 
with competing phases allowed by these symmetries. 

\subsection{Dependence on interaction strength}

\begin{figure}[t!]
\centerline{\includegraphics[height=7.0cm,angle=270]{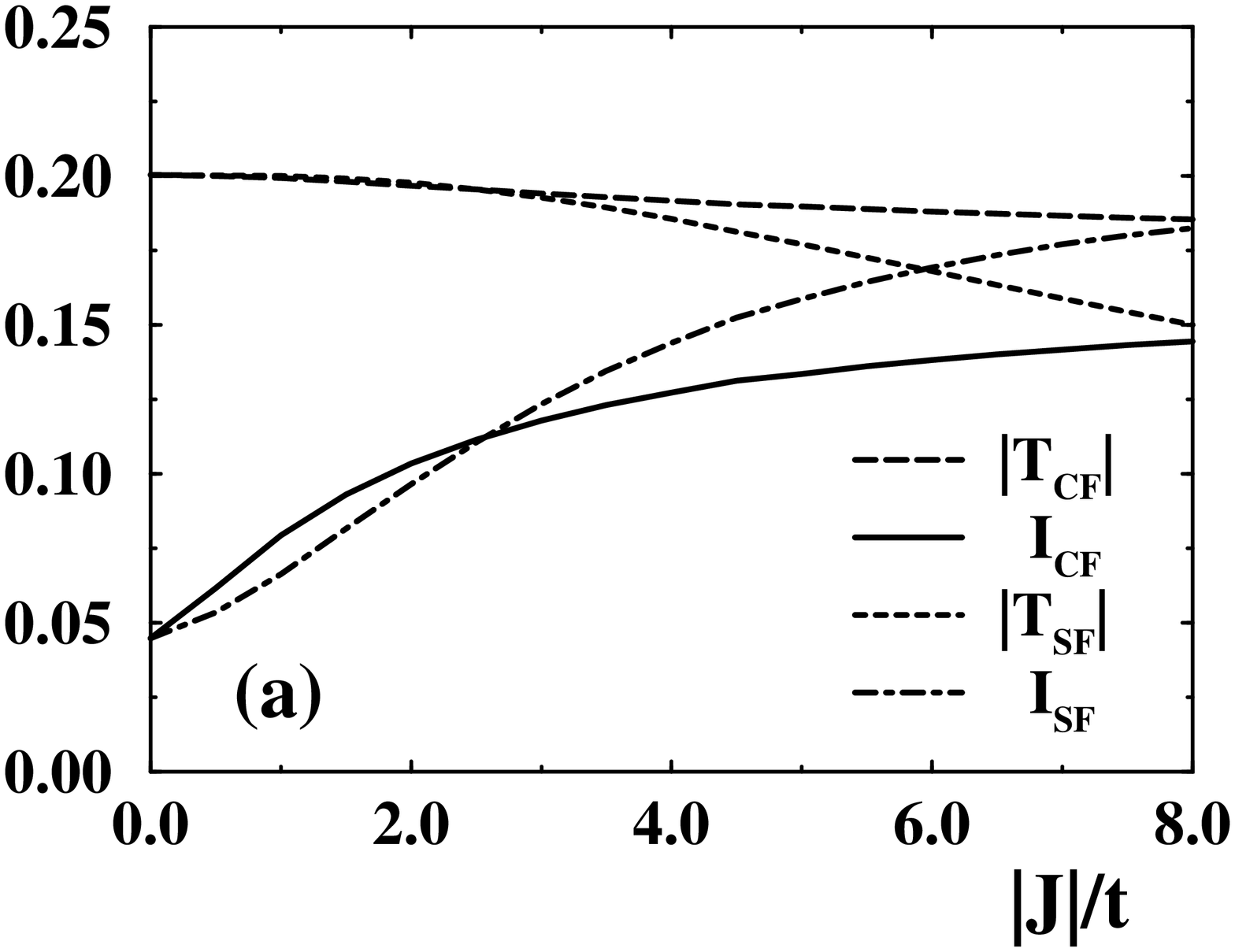}}
\centerline{\includegraphics[height=7.0cm,angle=270]{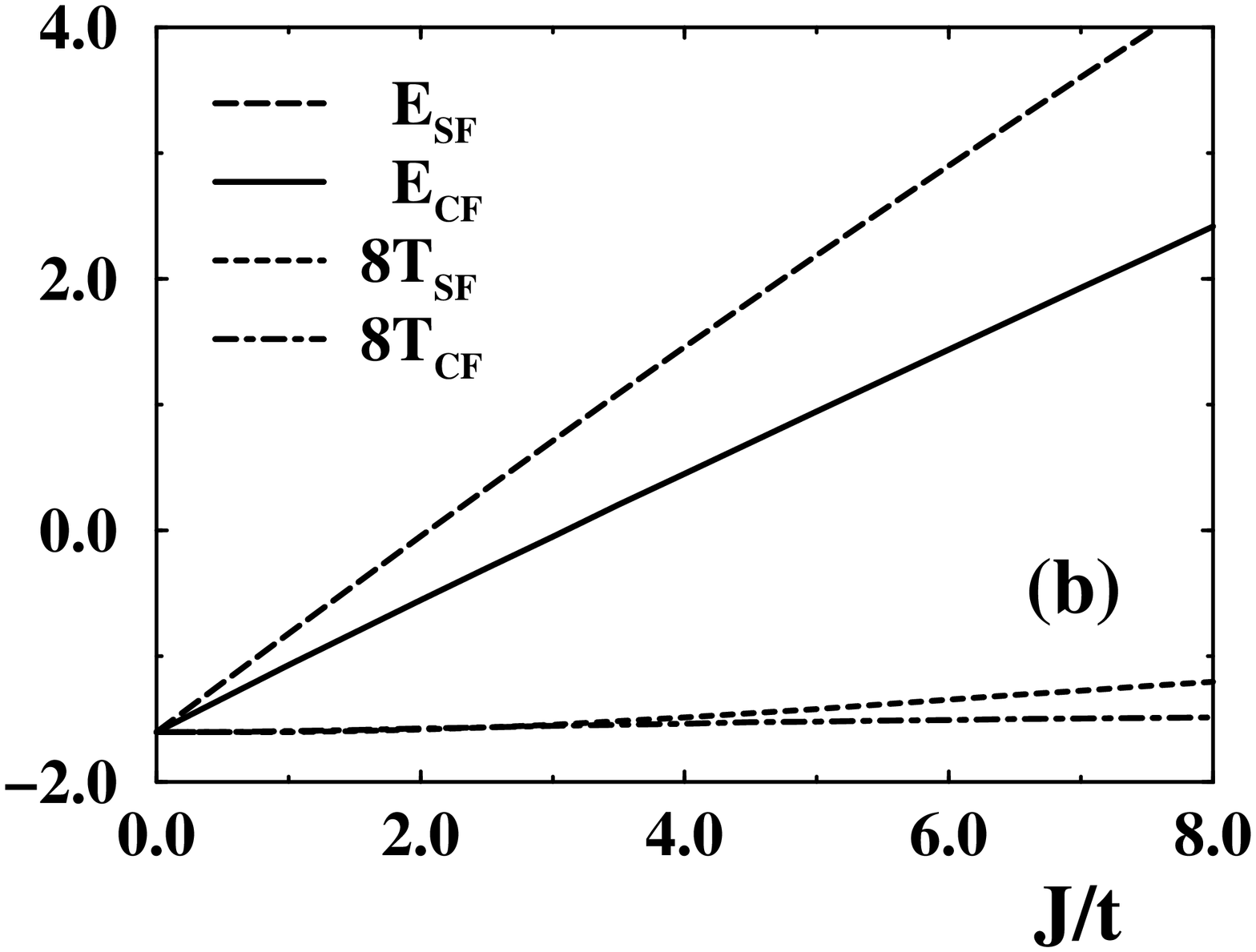}}
\caption{Properties of CF and SF phases with increasing interactions 
$U = |J| = 4V$ for doping $x$ = 0.028 as obtained in HFA: 
(a) kinetic energy $T$ and current $I$ per bond for each spin orientation; 
(b) total energy $E$ and kinetic energy $8T$. }
\label{xicer}
\end{figure}

The CF and SF states for small doping are characterized in 
Fig.~\ref{xicer}(a), which shows the quantities $T$ and $I$ as functions 
of interaction strength for a fixed choice of the interaction ratios 
$|J|/U = 4V/U = 1$. Qualitatively, the current does not contribute to 
the overall kinetic energy, and the finite imaginary part of the bond 
hopping parameter is stabilized by Fock terms originating from decouplings 
of the $J$ and $V$ interactions, as seen in the presence of the negative 
signs in the first line of Eq.~(\ref{edj}) and the second line of 
Eq.~(\ref{edv}). Figure \ref{xicer}(b) compares the kinetic energy 
contribution to the total energy, illustrating the dominance of the static, 
zeroth-order interaction terms contained in Eqs.~(\ref{edj}) and (\ref{edv}).
While the kinetic contribution remains nearly constant, the total energy 
increases monotonically, reflecting the penalty for failure to adopt 
charge- or magnetically ordered states. It is this failure which 
ultimately destabilizes the CF and SF phases, rendering them uncompetitive 
in comparative HFA studies in real space. 

One observes again that growth of the current has little effect on $T_{\rm 
CF}$ in Fig.~\ref{xicer}(a), and that $I_{\rm CF}$ saturates with increasing 
interaction strength, approaching the limit $|T_{\rm CF}| = |I_{\rm CF}|$ 
where the phase per bond is $\chi_{ij\sigma} = \pi/4$. Thus the phase $\phi$ 
for a plaquette approaches $\phi = \pi$, and the CF state is the $\pi$-flux 
phase of Ref.~\onlinecite{ram}. By contrast, in the SF state $I_{\rm SF}$ 
continues to rise with increasing interaction strength, the kinetic 
contribution $T_{\rm SF}$ falls, and this phase becomes unstable to a 
symmetry-broken magnetic state with finite net moment. We note also that 
the flux states remain stable in the limit of vanishing interactions, 
although in this case they are not the global energy minima of the 
system. A similar situation is found by analysis of different saddle 
points in a large-$N$ approach: for the SU($N$) saddle point, which 
yields the CF state, the circulating-current phase also survives at zero 
interaction strength. While Fig.~\ref{xicer} illustrates the situation for 
$x = 0.028$ (4 holes in a 12$\times$12 system), the same behavior is found 
at higher doping: for $x \simeq 0.1$ the evolution of the bond phase in the 
SF state is almost identical, while for the CF state it converges more 
slowly to $\pi/4$, and the current in the zero-interaction limit is 
somewhat smaller.

\subsection{Dependence on next-neighbor hopping}

\begin{figure}[t!]
\centerline{\includegraphics[height=7.0cm,angle=270]{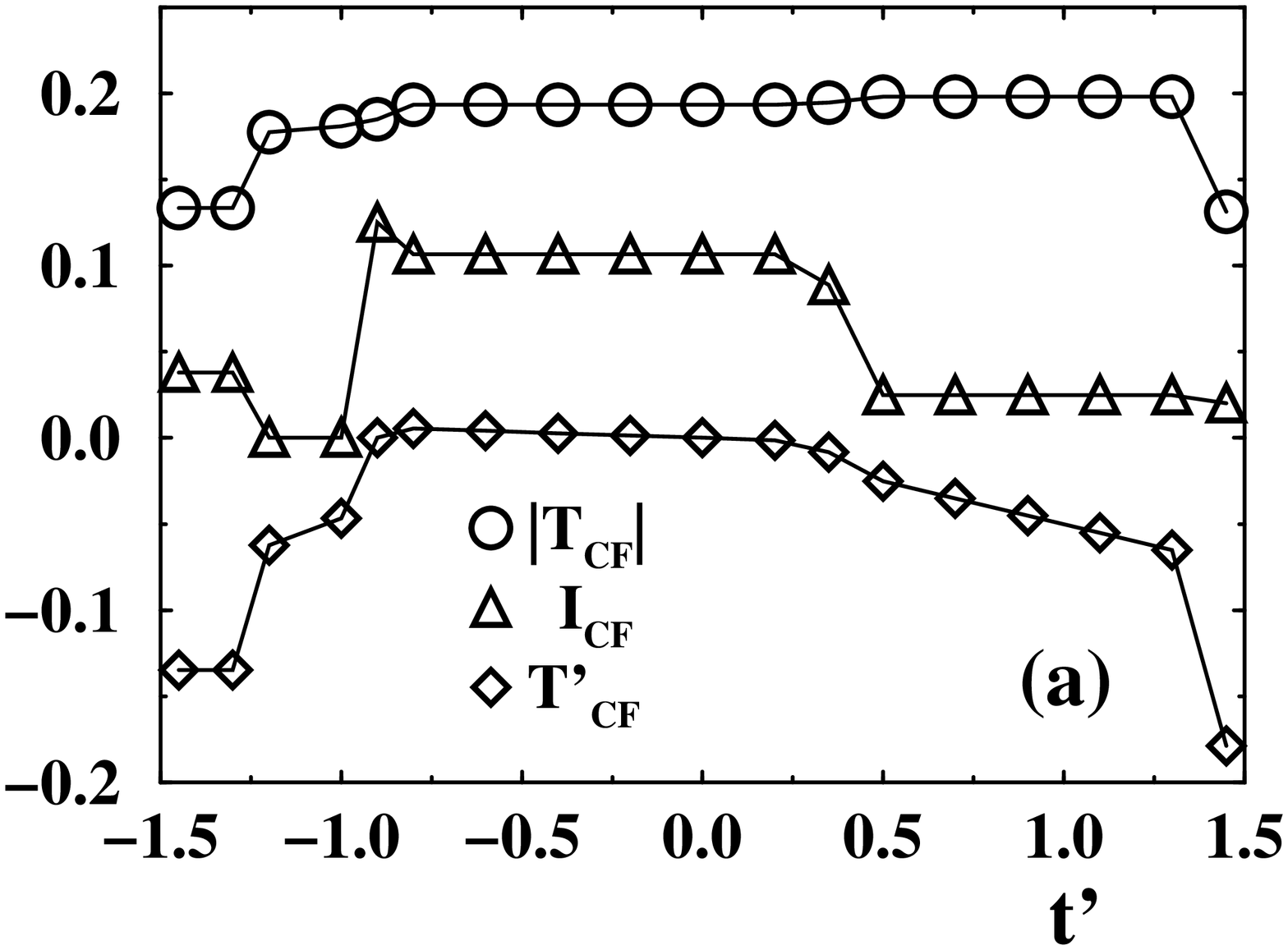}}
\centerline{\includegraphics[height=7.0cm,angle=270]{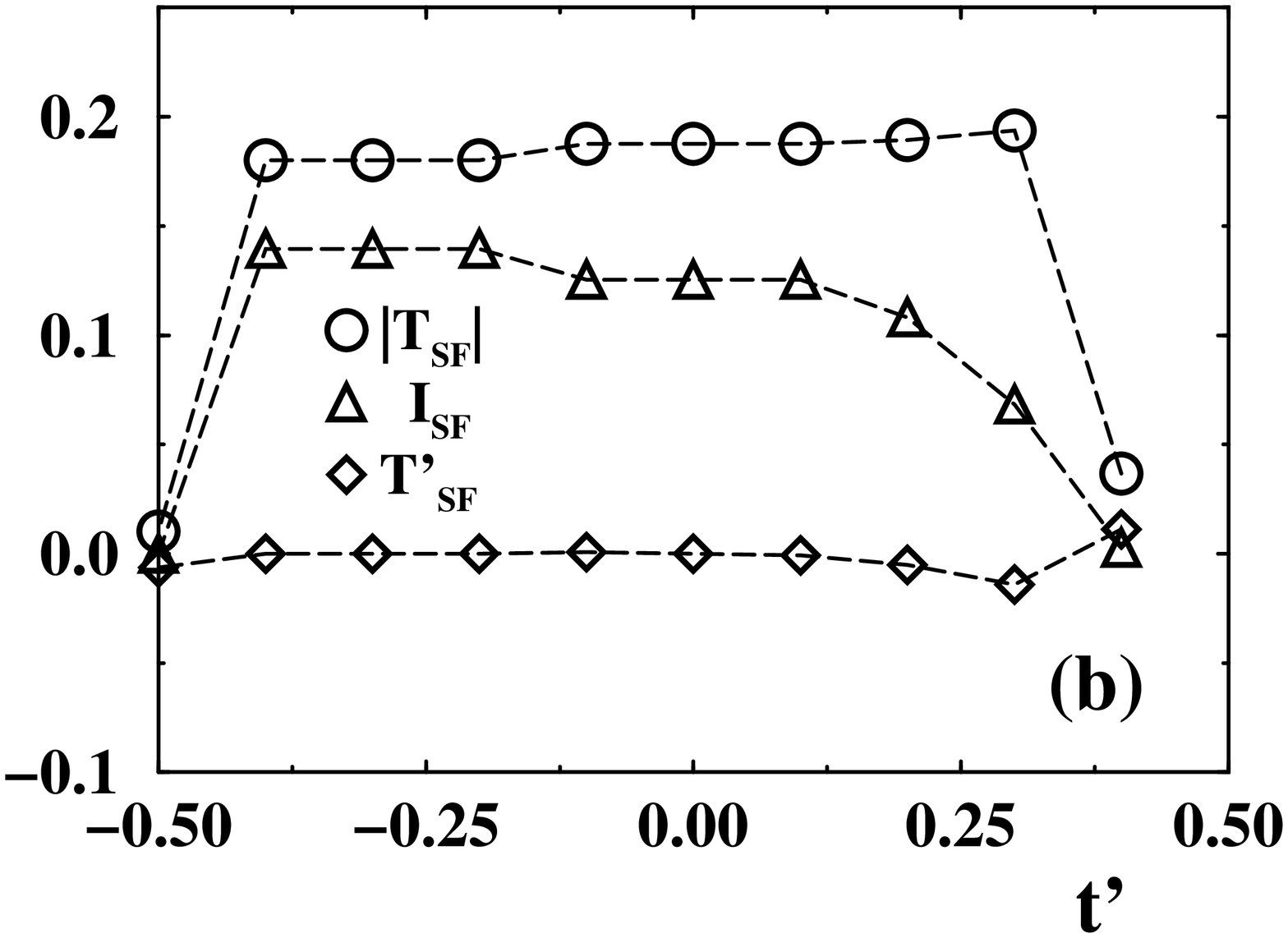}}
\caption{Influence of next-neighbor hopping $t'$ on flux phases with 
$U = |J| = 4V = 4t$ for doping $x$ = 0.083 as obtained in HFA: 
kinetic energy $T$ and current $I$ for each spin orientation per $x$- 
and $y$-bond, and kinetic energy $T'$ per ($x$+$y$)- and ($x$-$y$)-bond, 
are shown for (a) CF and (b) SF states.}
\label{xtpu}
\end{figure}

The effect of a next-neighbor, cross-plaquette hopping $t'$ in 
Eq.~(\ref{ehh}) is shown in Fig.~\ref{xtpu}. We have computed the 
bond-order parameters for both nearest- and next-nearest-neighbor bonds 
for charge- and spin-flux phases over the range of values of $t'$ for 
which these phases are maintained. The qualitative behavior is the 
same for all dopings in Fig.~\ref{yicr}, and is illustrated for an 
intermediate choice of $x$. We have defined $t$ and $t'$ with inclusion 
of a negative sign in Eq.~(\ref{ehh}), so $t = 1$ and $t' < 0$  
corresponds to the physical sign for cuprates. We note also that 
for clarity of presentation Fig.~\ref{xtpu} shows $|T| = - T$ 
but the true sign of $T'$. The most interesting feature is that 
small values of $t'$ have no effect at all on CF and SF states, in 
that the expectation values of the bond energy and current, which are 
determined largely by $H_J$ and $H_V$ in Eq.~(\ref{eehh}), are not 
changed by the additional kinetic term. In this regime, which persists 
more strongly for negative $t'$, diagonal hopping gives a very small 
contribution to the energy which is positive for $t' < 0$. Negative values 
of $t'$ show a small tendency to reinforce both types of flux phase, 
by which is meant to increase the bond current with a concomitant small 
decrease in bond energy, which in the context of cuprate systems is 
a noteworthy result. Positive values of $t'$ show a more pronounced 
tendency to suppress both types of flux phase; for the CF state there is 
a stable small-current regime over a range of values of $t'$, in which 
the next-neighbor hopping contribution to the kinetic energy is 
quite significant. 

The regions of stability are cut off for both signs of $t'$ as the need 
to minimize the total kinetic energy drives transitions to phases with 
similar energetic contributions from both nearest-neighbor and diagonal 
bonds. These states are marked by a sharp rise in $|T'|$. For the CF state 
this process occurs for $|t'| \sim t$, and involves homogeneous zero- or 
small-current phases. For the SF state the transitions involve inhomogeneous 
phases of small total kinetic energy and non-zero magnetic moments, and 
occur for significantly smaller $|t'|$. This latter result may be taken 
as evidence that the SF phase is generally less stable to competing phases 
[Fig.~\ref{xicer}(b)]. We stress that the values of $t'$ required to 
destabilize the CF and SF states at this filling should not be considered 
as quantitatively representative, and emphasize only the qualitative 
behavior in the presence of $t'$. The results are consistent with the 
renormalization-group analysis of Ref.~\onlinecite{rkk1} for the physical
sign of $t'$ ($t'< 0$ in our notation convention), in which the location of 
the CF phase boundary is scarcely affected for $|t'| < 0.3$, whereas the 
SF phase is entirely suppressed. Further studies of the role of diagonal 
hopping, including consideration of circulating-current phases established 
on the diagonal bonds at very large values of $|t'/t|$, have been conducted 
in Ref.~\onlinecite{rckk}.

We conclude this survey of the qualitative properties of flux phases 
in the weak-coupling regime by commenting that in the unrestricted 
Hartree-Fock approach essentially no inhomogeneous current-carrying 
states were found. While the method presents no barriers to phase 
coexistence, over most of the parameter range the coexisting flux-phase 
components are small compared to other, competing phases, and only a very 
weak, uniform current pattern is maintained. This situation may be altered 
in the presence of a relatively large next-neighbor hopping parameter $t'$, 
which also raises the possibility of obtaining specific inhomogeneous 
current configurations,\cite{rckk} but this parameter regime is of little 
physical interest in the context of cuprates, and is not considered here.

\section{Flux Phases with Strong Correlations}

\subsection{Gutzwiller Approximation}

The analysis of the preceding section is limited to the regime of weak 
coupling. However, interactions in cuprates are strong, and in particular 
the large on-site repulsion $U$ causes a suppression of double site 
occupation due to electronic correlations, which is not taken into account 
in the unrestricted HFA. 
As a consequence, the parameter regime in which flux phases were found 
in Sec.~III has little relation to the model parameters appropriate 
for superconducting cuprates. In this section we use the Gutzwiller 
Approximation (GA), which includes strong correlations through their 
effect in suppressing states including doubly occupied sites from the 
Hilbert space, and thus allows us to approach the cuprate parameter regime. 
Because the GA reproduces qualitatively the effects of (partially or fully) 
projected hopping on the single-particle level, an adapted HFA of the 
type employed in Sec.~III remains valid. Recent considerations of 
Gutzwiller wave functions for multiband models have also led to an 
improved understanding of the Mott transition in these systems.\cite{rbwgaf}

While many of the early analyses of flux phases in cuprate-related models 
employed an effective large-$U$ description such as the $t$-$J$ model, 
relatively few of these employed the projection of the hopping term 
associated with the constraint of suppressed double site occupancy. This 
constraint, which with full projection yields a hopping term of the form 
\begin{eqnarray}
H_t \! & = & \! - t \! \sum_{\langle i,j \rangle \sigma } [( 1 \! - \! 
n_{i\bar{\sigma}}) c_{i\sigma}^{\dag} c_{j\sigma} (1 \! - \! 
n_{j\bar{\sigma}}) + {\rm H. c.} ] \nonumber \\
&- & \! t' \! \sum_{\langle\langle i,j \rangle\rangle\sigma } \! 
[( 1 \! - \! n_{i\bar{\sigma}}) c_{i\sigma}^{\dag} c_{j\sigma} (1 \! 
- \! n_{j\bar{\sigma}}) + {\rm H. c.} ], 
\label{eph}
\end{eqnarray}
is technically very difficult to implement systematically. Because of its 
singular nature the connection between unprojected (independent electron)
and projected models remains quite unclear, and the possibility arises that 
flux phases may indeed be far more relevant for the latter. One exact means 
of implementing projected hopping is in numerical studies, such as the exact 
diagonalization analysis of uniform and staggered flux phases in the 
$t$-$J$ model\cite{rphr} and of the CF phase in the presence of hole 
pairing.\cite{rl,rwe} 

Here we approximate the projected model by the Gutzwiller approach,\cite{rg} 
in which electron correlation effects are contained within multiplicative 
prefactors modifying the terms of the Hamiltonian. These prefactors are 
obtained from statistical considerations based on the numbers of states 
available and excluded by the constraint.\cite{rv} This approximation to 
the effects of a large $U$ term was presented for the 2D Hubbard model in 
Ref.~\onlinecite{rzgrs}, where the technical details are also summarized. 
For the purposes of controling the extent of projection, here we 
apply these factors in their general form at finite $U$ (see also 
Ref.~\onlinecite{rz})
\begin{eqnarray}
\label{egf}
g_t & = & \frac{(n - 2 d)(\sqrt{d} + \sqrt{1 - n + d})^2}{n (1 - 
{\textstyle \frac{1}{2}} n)},                             \nonumber \\ 
g_J & = & \left[\frac{(n - 2d)} {n (1 - {\textstyle\frac{1}{2}} n)} 
\right]^2 \;\; = \;\; g_V^{11},  \nonumber \\ 
g_V^{12} & = & \frac{2 d (n - 2 d)}{n^3 (1 - {\textstyle\frac{1}{2}} 
n)}, \;\;\;\; g_V^{22} \;\; = \;\; \frac{16 d^2}{n^4} ,
\end{eqnarray}
where $g_t$, $g_J$, and $g_V^{m m'}$ multiply the corresponding terms in 
Eq.~(\ref{eehh}). The variable $n$ is the electron filling (Sec.~III.A), 
and the parameter $d$ represents the density of doubly occupied sites in 
the system, $d = \langle n_{i\uparrow}n_{i\downarrow}\rangle$. Because we 
consider only those uniform phases found in the HFA analysis of Sec.~III, 
the density $d$ is independent of $i$. For finite values of $U$, three 
terms are required to represent the nearest-neighbor Coulomb interaction, 
depending on the occupation state of the sites involved. Because the number 
of bonds with at least one doubly and one singly occupied site is low for 
all parameters and dopings under consideration, it is a very small and thus 
acceptable approximation to proceed with a single factor $g_V \equiv 
g_V^{11}$.
 
The effective model with projected hopping is considered to be derived from 
a Hubbard model with large but finite $U$.\cite{rtmgy} Thus the parameter 
$d$ is nonzero, in accord with the finite values of the $J$ and $V$ terms, 
whose physical origin lies in virtual hopping processes involving an 
effective double site occupancy. We define the quantity $d_0 (n)$ as the 
average value of the double site occupancy for a fixed, finite value of 
$U$ at each electron filling $n$, as determined in the HFA to the 
correlated state ($d_0(n) < (n/2)^2$ because this reference Hartree-Fock 
state is magnetic). An effective $t$-$J$-$V$  Hamiltonian with 
projection factors $g_t$, $g_J$, and $g_V$ determined using $d_0(n)$ in 
Eq.~(\ref{egf}) should then represent the starting $t$-$U$ Hamiltonian. 
Within this framework, projection to a lower net double occupancy $d < 
d_0 (n)$ is a variationally motivated procedure which seeks to minimize 
the total energy by further elimination of the energetically costly 
doubly occupied sites, the limit of this process being a full projection 
with $d = 0$. In Sec.~IV.C we will use $d$ as a variable parameter to 
illustrate the interpolation between the unrestricted HFA and the 
strongly correlated state for the same electron density $n$. We note,
however, that only small corrections to the energy and magnetization
are expected due to this procedure in the strongly correlated regime,
because the unrestricted HFA already captures most of the relevant 
effects, and for large $U$ approaches the same limit as the projected 
model.\cite{roz} While this partial projection procedure 
is of limited meaning in the context of a model with arbitrarily chosen 
values of $J$ and $V$ as in Sec.~III, the effective model used in this 
section is that obtained by large-$U$ expansion of the Hubbard model 
(\ref{ehh}). The parameters $J$ and $V$ in Eq.~(\ref{eehh}) thus have 
the specific values $J = 4 t^2 / U$ and $V = - J/4$. 

Further insight into the physical meaning of the GA and its relation to 
the Gutzwiller projection for cuprate systems, may be obtained from the 
consideration of variational wave functions, a study which has been 
undertaken both for superconducting states\cite{rprt} and for coexisting 
superconducting and CF states.\cite{ril} We note here that the partially 
projected states used in this framework are not appropriate for addressing 
the Mott transition at half-filling, as has been shown systematically from 
their violation of optical sum rules.\cite{rmc} At finite doping these 
considerations bear a certain similarity to the recent discussion of 
``gossamer'' superconducting states.\cite{rbls,rz}

For the purposes of analyzing flux phases in a cuprate-relevant model 
with projected hopping, we begin by considering a Hubbard model with 
$U$ = 12. The Hartree-Fock density $d_0 (n)$ of doubly occupied sites 
is then computed for all values of the filling $ 0.8 \le n \le 1$ ($0 
\le x \le 0.2$). An effective projected model is obtained by computing 
the Gutzwiller factors [Eq.~(\ref{egf})] at each filling by varying $d$  
between the values $d = 0$, which describes full projection and is 
implemented to extract the properties of correlated electrons as 
described by a $t$-$J$ model, and $d = d_0 (n)$, which should reproduce 
qualitatively the unprojected (Hartree-Fock) states. The final 
version of the effective model is then as in Eq.~(\ref{eehh}), but with 
$t$ replaced by $g_t t$, $J$ replaced by $g_J t/3$, $V$ by $- g_V t/12$, 
and no on-site repulsive term. We emphasize that the derivation of this 
effective model involves a singular truncation rather than a systematic 
transformation, with the consequence that no meaningful energetic 
comparisons are possible between the results of Sec.~III and the results 
obtained after projection. However, it is important to understand the 
general properties of the projected system, and comparisons performed 
within it remain well defined.

\subsection{Qualitative consequences of Gutzwiller Approximation}

Qualitatively, one of the principal effects of projection is to sharpen 
the competition between the different candidate ground states. The 
exclusion of doubly occupied states eliminates most of the possibilities 
for a coexistence of localized and delocalized phases, and phase boundaries 
would thus be expected to become better defined. For CF phases these 
boundaries may also move closer to a parameter regime meaningful for cuprates 
(see below): when viewed as a means for dynamical holes to avoid each 
other (by circulating around each other on plaquettes), the CF state 
indeed allows a greater spatial separation than the superconducting state 
while retaining a favorable kinetic energy. However, application of this 
schematic picture for the SF phase, which requires in principle that 
up- and down-spin particles can move past each other on each individual 
plaquette, suggests that this type of state would be completely 
suppressed by projection. 

By inspection of the Gutzwiller prefactors (\ref{egf}) it is clear that close 
to half-filling the properties of the projected model are significantly 
different from those of the unprojected model. With full projection the 
factor $g_t$ vanishes as $x \rightarrow 0$, giving a complete suppression 
of the kinetic contributions to the total energy, and also of any 
bond currents. This suppression is large for small doping, where the 
kinetic and current terms are significantly smaller than in the 
unprojected system. However, because all competing states are similarly 
affected by projection, this does not imply that circulating-current states 
must become less competitive. By contrast, for small $d$ the prefactor 
$g_J \rightarrow 4$ as $x \rightarrow 0$, suggesting a relatively strong 
enhancement of magnetic interactions. Once again, because the superexchange 
interaction has both spin and bond-order decouplings [see Eq.~(\ref{edj})], 
this trend does not necessarily imply a suppression of flux states. 
The prefactor $g_V$ is smaller than $g_J$, and is not expected to play a 
significant role. For larger values of $x$ these statements remain generally 
true while $U$ is large, and $d$ correspondingly small, but the relative 
strengths of the kinetic and magnetic terms will change almost linearly 
with $x$.

\begin{figure}[t!]
\centerline{\includegraphics[height=7.0cm,angle=270]{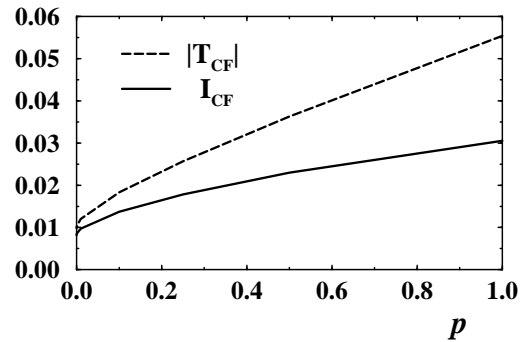}}
\caption{Kinetic energy $T$ and current $I$ per bond for each spin 
orientation in a CF state for with partially projected hopping for 
$x$ = 0.028 and $J = - 4V = t/3$.}
\label{yicsfr}
\end{figure}

\subsection{Partial projection}

We begin the more quantitative part of the analysis by considering the 
evolution of the flux phases with the extent of projection of the hopping 
term. Following the above discussion we define a partial projection variable 
\begin{equation}
p = d / d_0 (n), 
\label{pro}
\end{equation}
such that $p = 0$ corresponds to fully projected hopping (\ref{eph}), 
and $p = 1$ to a projection at finite $U$ which represents (by reference 
to the average number of doubly occupied sites) the HFA of Sec.~III. 
Fig.~\ref{yicsfr} shows the bond energy and current of a CF 
state as a function of $p$ for a system with 140 electrons ($x$ = 
0.028); while this doping illustrates most clearly the effects of 
projection, qualitatively identical results are found for higher values. 
At larger values of $p$, which correspond to relaxation of the constraint 
and thus to a larger relative kinetic contribution, $|T_{\rm CF}|$ 
and $I_{\rm CF}$ change linearly with $p$, and have a fixed ratio 
(corresponding to a plaquette phase $\phi\sim 2 \pi/3$) which decreases 
with increasing doping. As full projection ($p = 0$) is approached the 
suppression becomes stronger, and the real and imaginary parts tend to 
fixed values determined by the filling; their ratio increases with 
projection up to an equivalent flux $\phi = 0.88\pi$ for this doping. 

The SF phase exists only for negative values of the superexchange 
$J$, which is not a valid parameter regime in the framework of a small 
$(t/U)$ expansion for the Hubbard model. If the sign of $J$ is simply 
inverted, the calculation does not converge to a stable local 
minimum corresponding to a SF state for any partial projection. The 
projection factors contribute in two ways to this result: the fact that 
$I_{\rm SF}$ approaches $|T_{\rm SF}|$ ($\phi \rightarrow \pi$) reduces 
a difference term [which for the CF state is a sum, causing the difference 
visible in Fig.~\ref{xicer}(b)] contributing a negative energy, and the 
suppression of the total kinetic energy removes what is then an essential 
negative contribution. This failure to form an SF state in the projected 
real-space HFA, which extends to all values of the doping $x$, may be 
taken as evidence that the schematic picture of counter-flowing up- and 
down-spin currents, which implies that electrons must hop past each other 
on each site, is in fact a valid description of this phase.  

With regard to the total energy of partially projected CF states, the 
increasing kinetic contribution as projection is relaxed ($p \rightarrow 
1$) means that the fully projected state has the highest energy. We have 
commented in Secs.~II and IV.A on the restrictions concerning meaningful 
energetic comparisons because of both the nature of the HFA and the 
singular projection scheme. Within the manifold of projected CF states, 
changes in total energy as a function of $p$ are dominated by the kinetic 
energy term, and, in the framework of a consistent derivation from the 
Hubbard model, no systematic study of energies, for example as a function 
of interaction strength [Fig.~\ref{xicer}(b)], is then possible. However, 
it is clear from our calculations in the projected regime that the energy 
difference between the CF state and other competing phases, such as 
antiferromagnetic order, is small. 

While we caution as before that quantitative conclusions cannot be 
drawn, and note also that energy scales are altered by projection, it 
does appear meaningful to state that, unlike in the unprojected framework, 
CF phases are strongly competitive in a projected model. This statement 
applies not only to the most favorable choice of interactions (Sec.~III) 
but specifically for cuprate parameters. The enhanced competitiveness of 
the CF state indicates that projection provides a better reflection of 
the energy gained by dynamically stabilized phases, the origin of which 
lies primarily in bond-order decouplings of the $J$ and $V$ terms. We may 
conclude from this consideration of the GA that CF phases are not in fact 
unrealistic for cuprate interactions when appropriate projection of the 
hopping terms is performed. More quantitative analyses of this issue are 
required. 

\subsection{Dependence on doping}

\begin{figure}[t!]
\centerline{\includegraphics[height=7.0cm,angle=270]{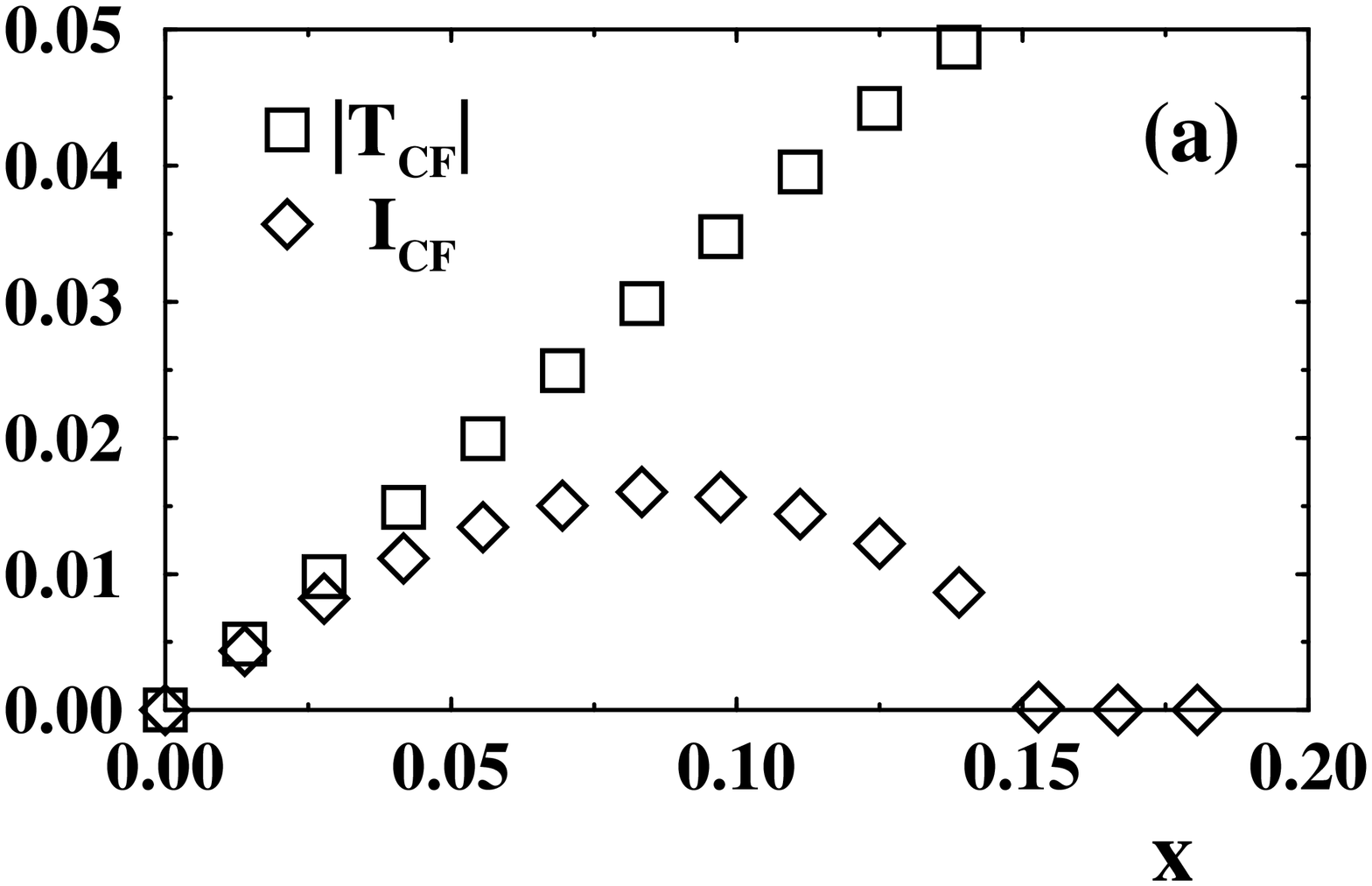}}
\centerline{\includegraphics[height=7.0cm,angle=270]{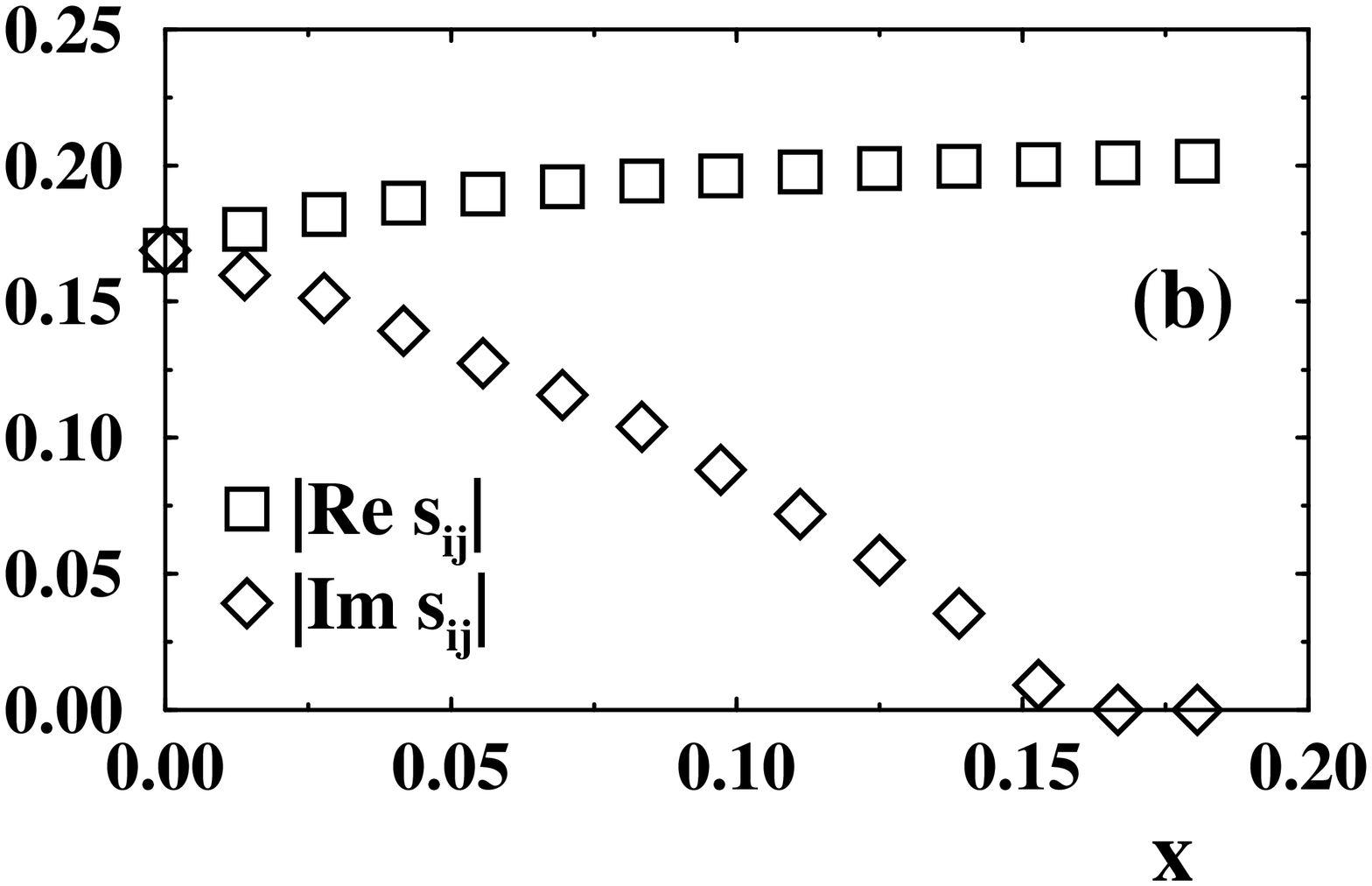}}
\caption{(a) Doping-dependence of kinetic energy $T$ and current $I$ per 
bond for each spin orientation in the fully projected ($p = 0$) CF phase 
with $J = - 4V = t/3$. (b) Corresponding bond-order parameters 
${\rm Re}\; s_{ij}$ and ${\rm Im}\; s_{ij}$. }
\label{yicpr}
\end{figure}

We consider next the dependence of the fully projected ($p = 0$) CF
state on doping. The evolution of the bond parameters $|T_{\rm CF}|$ and 
$I_{\rm CF}$ with $x$, shown in Fig.~\ref{yicpr}(a), exhibits three 
important qualitative features. First, the effective projected model 
approaches the $\pi$-flux phase, $|T_{\rm CF}| = I_{\rm CF}$, when the 
doping decreases towards half-filling. However, in this limit the kinetic 
energy and current also vanish in the projected model, in contrast to the 
unprojected one. Secondly, the kinetic component increases linearly 
in magnitude (becomes more negative) with hole doping across the full 
range of $x$, tending towards the value found for a projected metallic 
state (not shown). Finally, the bond current grows at first, but beyond 
$x\sim 0.08$ falls continuously to zero (at $x_c \simeq 0.16$), indicating 
the instability of the CF state to a metallic (Fermi-liquid) phase at higher 
doping. 

We stress that in the projected model the kinetic terms are no longer 
identical to the real and imaginary parts of the bond-order parameter 
$s_{ij}$, but are related by a ratio of Gutzwiller factors (\ref{egf}). 
The components of $s_{ij}$, which are determined by the bond decoupling 
of the terms $H_J$ and $H_V$ in Eq.~(\ref{eehh}), are shown as functions 
of doping in Fig.~\ref{yicpr}(b). The absolute value of the bond amplitude 
$|s_{ij}|$ decreases only slightly with doping, while the bond phase 
falls continuously from $\pi/4$ at half-filling to zero at $x_c$. At a 
qualitative level, the monotonically decreasing ${\rm Im}\; s_{ij}$ and 
dome-shaped form of $I_{\rm CF}$ are rather similar to the behavior of the 
pseudogap amplitude and order parameter for $d$-wave superconductivity 
computed in the projected BCS wavefunction framework in 
Ref.~\onlinecite{rprt}. Given that we have chosen to allow only bond 
order, but no pairing decouplings, such a similarity may be expected 
from the SU(2) symmetry relationship of CF and superconducting phases 
in this class of models.\cite{rlnnw} The microscopic origin of a 
coexistence of paired and CF states has been investigated by a number 
of authors.\cite{rilw,rl,rwe} At the quantitative level, the net hopping  
kinetic energy per site $|E_t| = 8|T_{\rm CF}|$ in Fig.~6(a) is for all 
dopings within 15\% of the value calculated in Fig.~4(a) of 
Ref.~\onlinecite{ril} using a projected wave function approach to the 
SC phase. This result underlines the energetic similarity of the SC 
and CF phases, and the need for a detailed and unbiased analysis to 
distinguish between them. 

We discuss only briefly the dependence of the CF phase on the strength 
of electronic interactions in the extended Hubbard model. Within the 
framework we have established for the Gutzwiller analysis this means 
an alteration in the value of $U$, and for increasing $U$ one approaches 
the asymptotic regime where $d_0 (n) \propto (t/U)^2$. However, as 
$d_0 (n)$ becomes smaller than its value for $U \simeq 8$, the Gutzwiller 
projection factors (\ref{egf}) are governed primarily by the doping 
$x$. This reflects the fact that increasing doping causes a gradual 
release of the constraints arising from strong correlations. In the 
large-$U$ regime of most interest it is thus the filling rather than 
the double occupancy which is the most important parameter determining 
the current and the bond phase of the CF state. 

\subsection{Dependence on next-neighbor hopping}

\begin{figure}[t!]
\centerline{\includegraphics[height=7.0cm,angle=270]{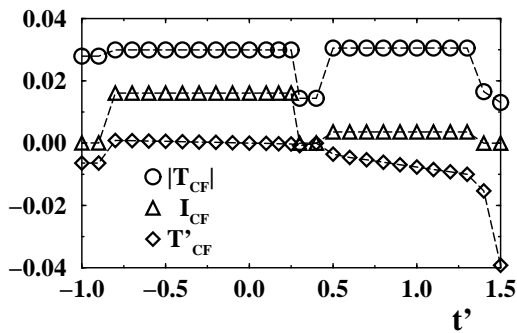}}
\caption{Influence of next-neighbor hopping $t'$ on the projected CF phase 
with doping $x$ = 0.083, showing kinetic energy $T_{\rm CF}$ and current 
$I_{\rm CF}$ per spin for each $x$- and $y$-bond, and kinetic energy 
$T'_{\rm CF}$ per ($x$+$y$)- and ($x$-$y$)-bond.}
\label{xtpp}
\end{figure}

Following the analysis of Sec.~III, we consider also the effect of 
a next-neighbor hopping $t'$ on the fully projected CF phase, shown 
in Fig.~\ref{xtpp}. As observed also for the unprojected model 
(Fig.~\ref{xtpu}), small values of $t'$ have no effect on the bond-order 
parameters $T$ and $I$, and the $T'$ contribution, which is positive at 
$t' < 0$, is essentially a residual consequence of the prevailing CF 
state. The regime of stability for the projected CF phase extends 
considerably further for negative than for positive $t'$, with its 
midpoint at $t' \sim - 0.3$: it appears reasonable to state that for 
the physical magnitude and sign of $t'$ in cuprates, next-neighbor 
hopping does not disrupt and may in fact reinforce the CF state. 

The CF phase is again terminated by abrupt transitions to zero-current 
states for sufficiently large values of $t'$. For $t' > 0$ a transition 
occurs at rather small $t'$ to a phase with strongly reduced bond kinetic 
energy. However, beyond this we find a further CF phase with a small but 
finite current, which permits a significant $T'$ contribution. This new 
phase persists to $t'/t > 1$ before being replaced by another state, of 
zero current, which again has strongly reduced $|T|$ due to manifestly 
frustrated hopping. For $t' < 0$ the conventional CF phase persists 
down to $t'/t \sim - 1$ before being replaced by a zero-current state 
which achieves negative kinetic energy on both nearest-neighbor and 
diagonal bonds,\cite{rnk2} with $|T|$ only slightly suppressed from 
its value in the CF state.

\section{Ring-Exchange Interaction}

We turn next to the effects of the four-spin ring-exchange interaction 
on the flux states of Secs.~III and IV. Because the structure of this 
term is quadratic in bilinear spin terms (\ref{ehk}), the structure of 
its Hartree-Fock decomposition (\ref{edk}) is similar to that of 
the superexchange term (\ref{edj}). However, in addition to 
nearest-neighbor spin and bond expectation values it contains also 
next-neighbor bond-order terms corresponding to the diagonals of each 
plaquette. These appear in the Hamiltonian matrix in the same position 
as next-neighbor ($t'$) hopping terms. The quartic nature of the 
ring-exchange interaction $K$, and the relatively small expectation 
values of the order parameters ($\langle {\bf S}_i {\bf \cdot S}_j 
\rangle \le 0.25, |s_{ij}| \le 0.25$), suggest that the effect of this 
term in an unprojected model will be small for all but the largest 
values of $K$. 

\begin{figure}[t!]
\centerline{\includegraphics[height=7.0cm,angle=270]{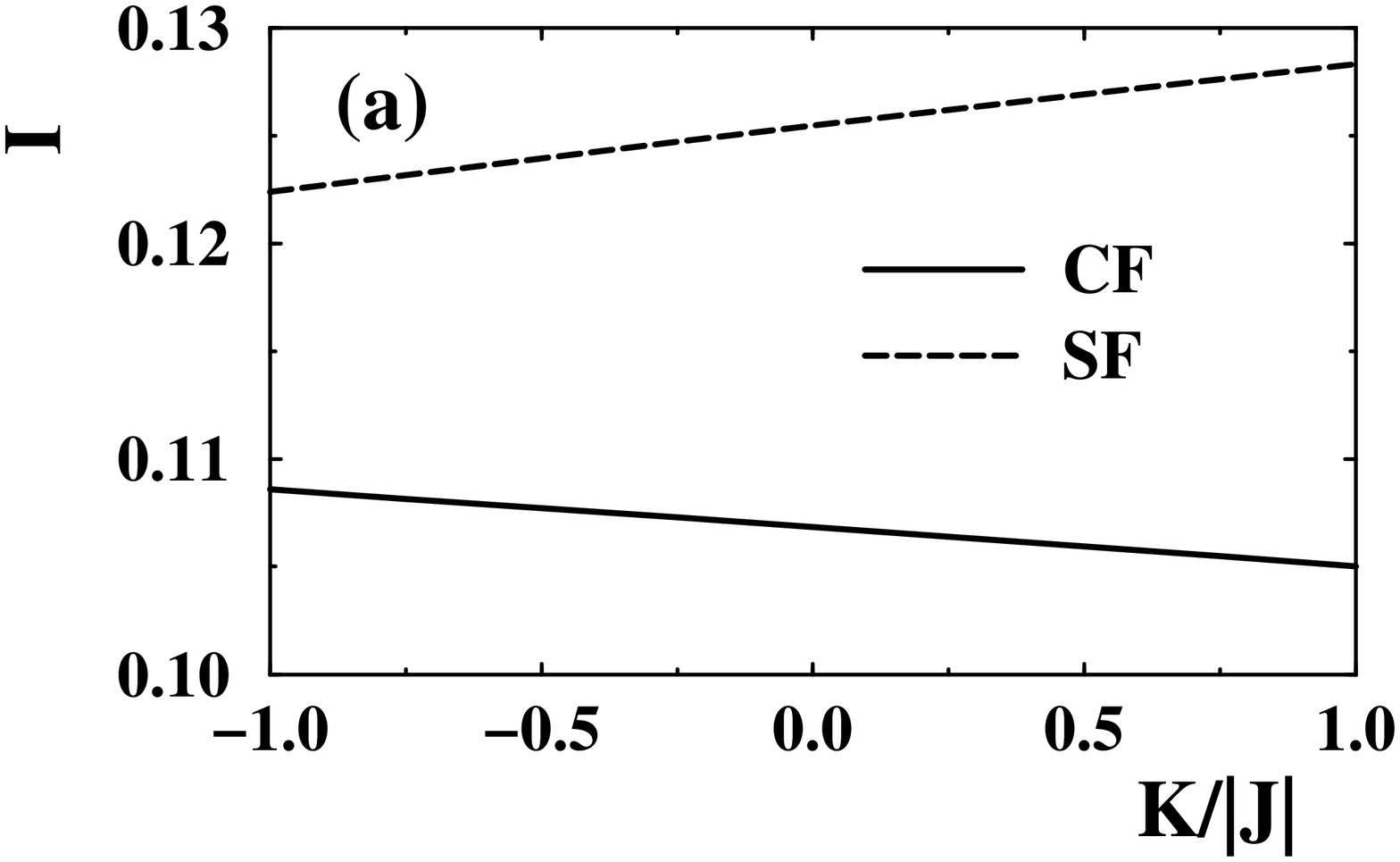}}
\vskip -.5cm
\centerline{\includegraphics[height=7.0cm,angle=270]{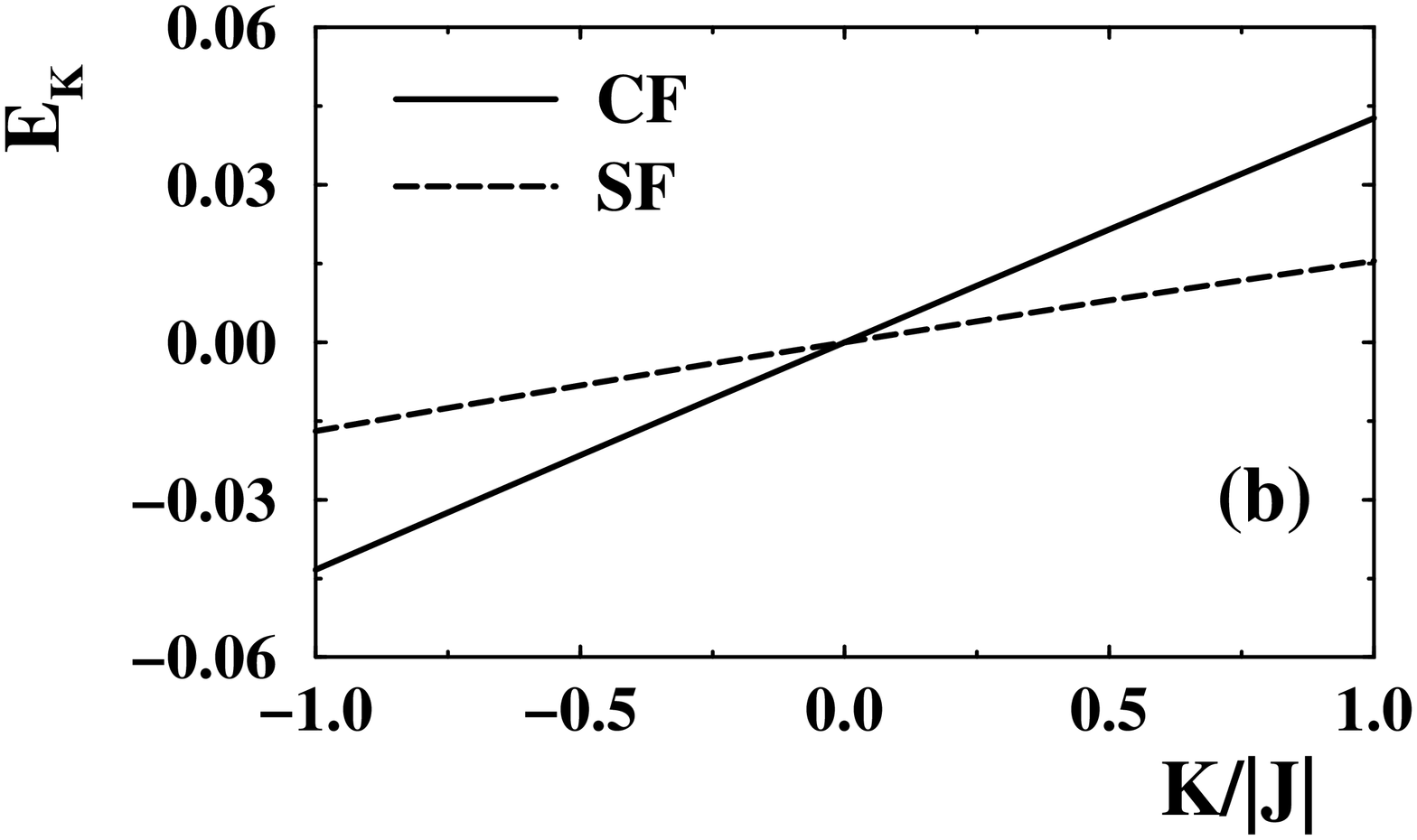}}
\caption{Effects of a ring-exchange interaction $K$ on unprojected CF 
and SF phases for $U = |J| = 4V = 4t$ at doping $x$ = 0.083: 
(a) current $I$ per bond per spin orientation; 
(b) energy contribution $E_{K}$. }
\label{xkcer}
\end{figure}

\subsection{Unprojected model}

The qualitative influence of the ring-exchange interaction on the 
unprojected CF and SF phases of Sec.~III may be assessed by considering 
the changes in bond currents, and also the total energy contribution, 
due to the presence of the four-spin coupling term. Figure \ref{xkcer}(a) 
shows the bond current in a system with a representative intermediate 
value of doping for a canonical choice of flux-phase interaction ratios 
$U = |J| = 4V = 4t$ over a range of values $-|J| \le K \le |J|$. It is 
clear that a positive ring-exchange interaction suppresses the CF state, 
whereas it acts to enhance the current in the SF state. 

These qualitative results may be explained by considering the 
``interference'' of currents on opposite bonds of each plaquette, 
which are coupled by the $K$ term in the manner shown in the first 
line of Eq.~(\ref{edk}), and specifically from the role in this 
interference of currents with opposing spins [second term of 
Eq.~(\ref{edsisj})]. The suppression of the CF state by a 
ring-exchange interaction is in agreement with the findings of 
Ref.~\onlinecite{rckk}; enhancement of the SF state is also consistent 
with a recent exact diagonalization study of a small cluster with $K$ 
as the dominant interaction, which revealed a type of SF phase.\cite{rlu} 
We comment further on this result below. 

The ring-exchange contribution to the total energy, shown in 
Fig.~\ref{xkcer}(b), is by its quartic nature always positive for $K > 0$.
This energy, $E_{K}$, constitutes a significantly smaller penalty in the 
SF state than in the CF state. Examination of the total energy (not shown) 
reveals a similar increase for the CF state, reinforcing the result of the 
preceding paragraph. For the SF state the total energy also shows a small 
increase of the same order as $E_{K}$, suggesting that changes in the 
kinetic and magnetic energies induced by the enhanced current do not 
compensate the positive ring-exchange contribution. As a function of 
filling, the influence of the $K$ term on the quantities characterizing 
the flux phases remains as shown in Fig.~\ref{xkcer} and displays no 
strong variation over the range $0 \le x \le 0.2$. 

\begin{figure}[t!]
\centerline{\includegraphics[height=7.0cm,angle=270]{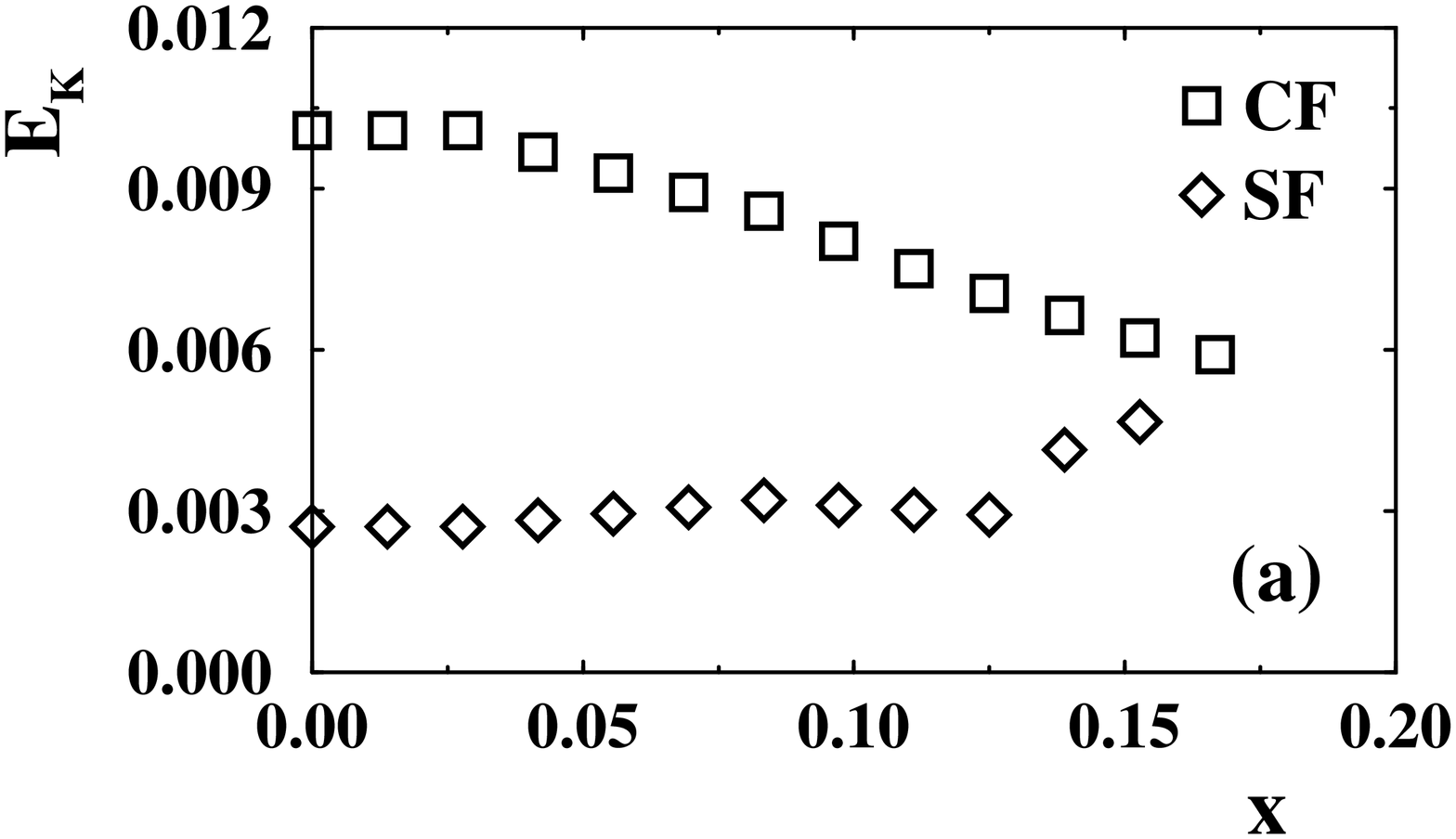}}
\vskip -.5cm
\centerline{\includegraphics[height=7.0cm,angle=270]{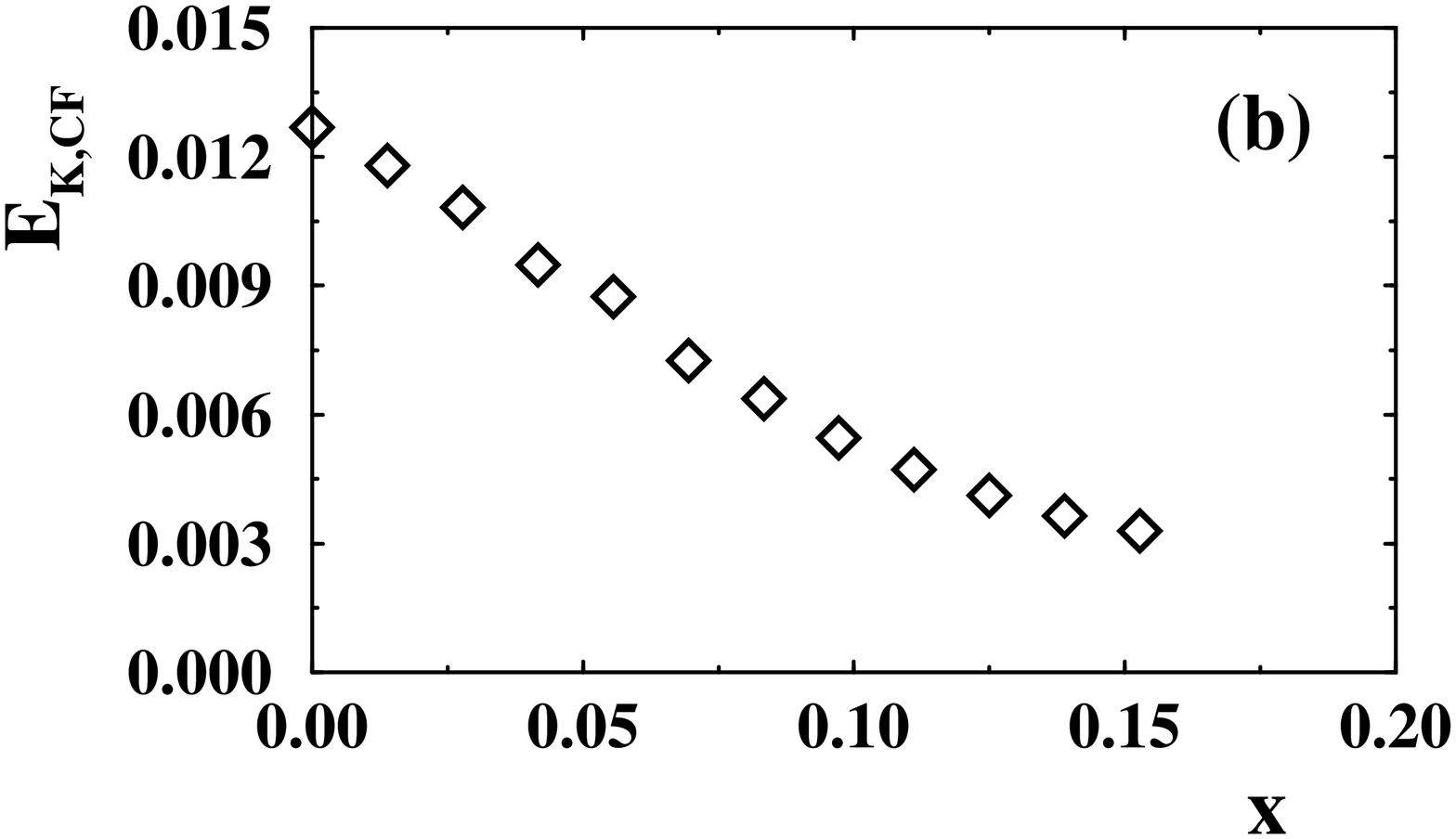}}
\caption{Doping-dependence of $E_{K}$ in the presence of a ring-exchange 
interaction $K = |J|/5$: (a) for unprojected CF and SF phases with 
$U = |J| = 4V = 4t$; (b) for the projected CF phase. }
\label{ykex}
\end{figure}

We have also computed the bond currents and ring-exchange energies of flux 
phases with $K=|J|/5$ for the full range of doping values over which these 
exist. The current differs only very slightly from its values at $K = 0$ 
(Fig.~\ref{yicr}), although there is a slight suppression which brings 
the CF transition to a lower doping value. The energy $E_{K}$, which 
is illustrated in Fig.~\ref{ykex}(a), shows only minor changes with 
increasing $x$ until the transition to a different (non-flux) state 
brings abrupt alterations at the critical dopings. The relative changes 
in current and energy are both understood readily from the evolution of 
$s_{ij}$ in the flux states at $K = 0$ (Sec.~III). However, we draw 
attention of the magnitude of these effects: even for $K = |J|$ these 
are in the percent range, and for the physical parameters of cuprate 
systems, $K \sim J/5$, they are quantitatively irrelevant. 

\begin{figure}[t!]
\centerline{\includegraphics[height=7.0cm,angle=270]{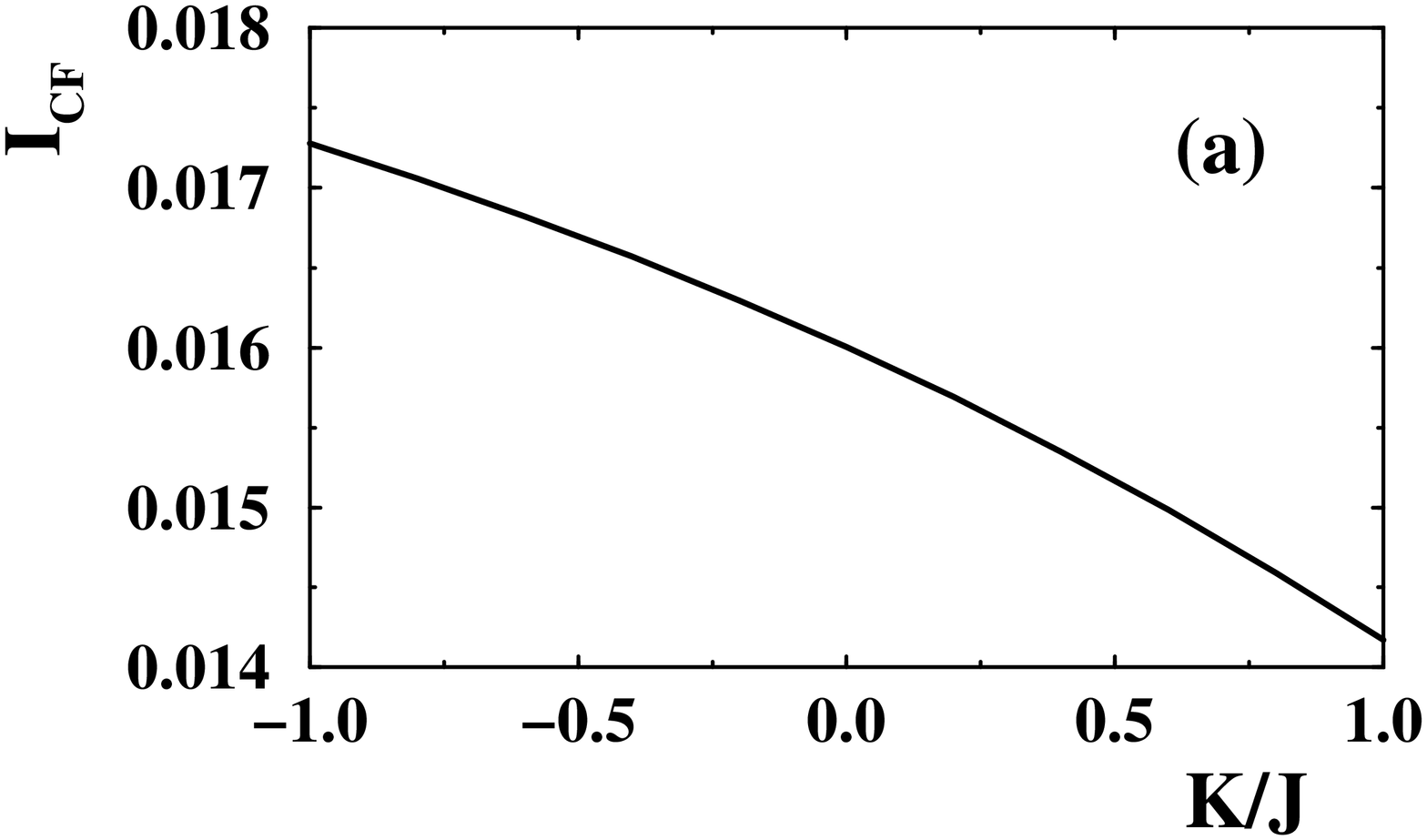}}
\vskip -.7cm
\centerline{\includegraphics[height=7.0cm,angle=270]{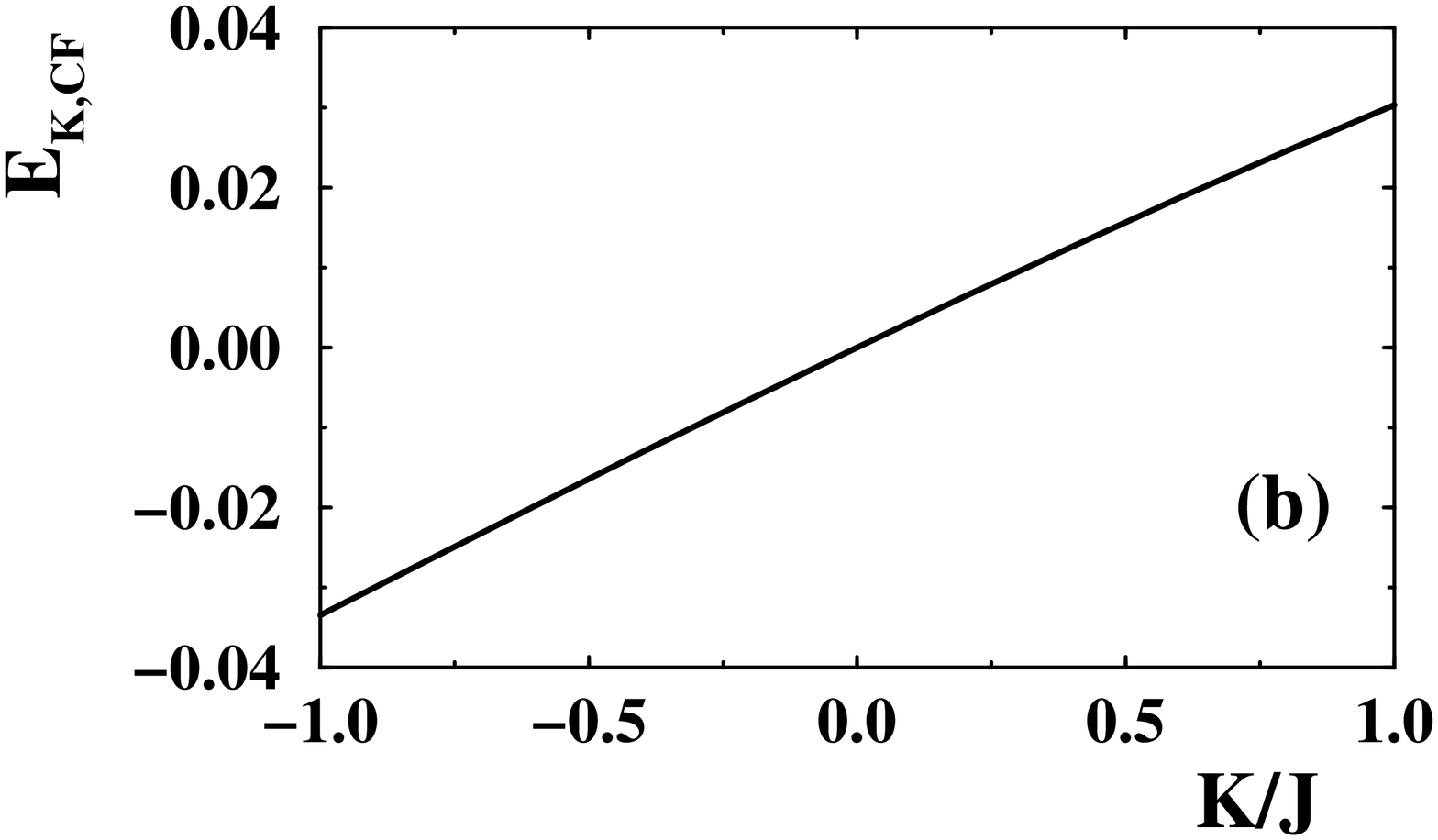}}
\caption{Effects of a ring-exchange interaction $K$ on the projected CF 
phase with $J = - 4V = t/3$ and doping $x = 0.083$: (a) current $I$ per 
bond per spin orientation; (b) energy contribution $E_{K,{\rm CF}}$. }
\label{ykcer}
\end{figure}

\subsection{Projected model}

In Sec.~IV we have shown that circulating-current states are 
significantly more competitive in projected models for strongly 
correlated electrons than in HFA. On first inspection one may also 
expect ring-exchange interactions to be considerably more relevant in 
a projected Hamiltonian: the appropriate Gutzwiller prefactor for the 
$K$ term is $g_K = g_J^2$, which approaches 16 at half-filling. To 
examine this possibility, in Fig.~\ref{ykcer} we show for an 
intermediate doping the projected analog of Fig.~\ref{xkcer}, following 
the bond current and ring-exchange energy in a fully projected CF state 
as a function of $K$. Both quantities show the same essentially linear 
trend which follows from suppression of the CF phase in the presence of 
a positive ring-exchange interaction. With regard to the 
doping-dependence of $I_{\rm CF}$ and $E_{K,{\rm CF}}$, as in the 
unprojected case the current is dictated by the $K = 0$ behavior of 
the system [Fig.~\ref{yicpr}(a)], with a small reduction in critical 
doping for positive $K$. The ring-exchange energy for the projected 
model with $K = J/5$ is shown in Fig.~\ref{ykex}(b), and exhibits a 
steady decline which may be traced to the combination of $s_{ij}$ 
parameters [Eqs.~(\ref{edk}, \ref{edsisj})] as the bond phase decreases 
from $\pi/4$ to 0 [Fig.~\ref{yicpr}(b)]. 

Returning to Fig.~\ref{ykcer}, in fact the ring-exchange effect on the 
current in the projected CF phase is significantly larger than in the 
unprojected one (10\% compared to 2\% at $K = J$ for $x = 0.083$). This 
result holds for all dopings, and extends also to the relative magnitude 
of the energy contribution $E_K$. However, it is clear that changes in 
both energy and current remain extremely small: this is not surprising 
given that the energetic contribution of the $K$ term is multiplied by 
the fourth power of bond expectation values $s_{ij}$ which are suppressed 
towards zero at half-filling for full projection. In this context we note 
that ordered magnetic states suffer no such suppression of the order 
parameter on projection, and thus from the large value of $g_K$ a 
Gutzwiller-projected model near half-filling might be expected to favor 
the formation of magnetic phases stabilized by large values of $K$. 

\subsection{Phases in the large-\mbox{${\bm K}$} regime}

\begin{figure}[t!]
\mbox{\hspace{0.3cm}\includegraphics[width=3.6cm]{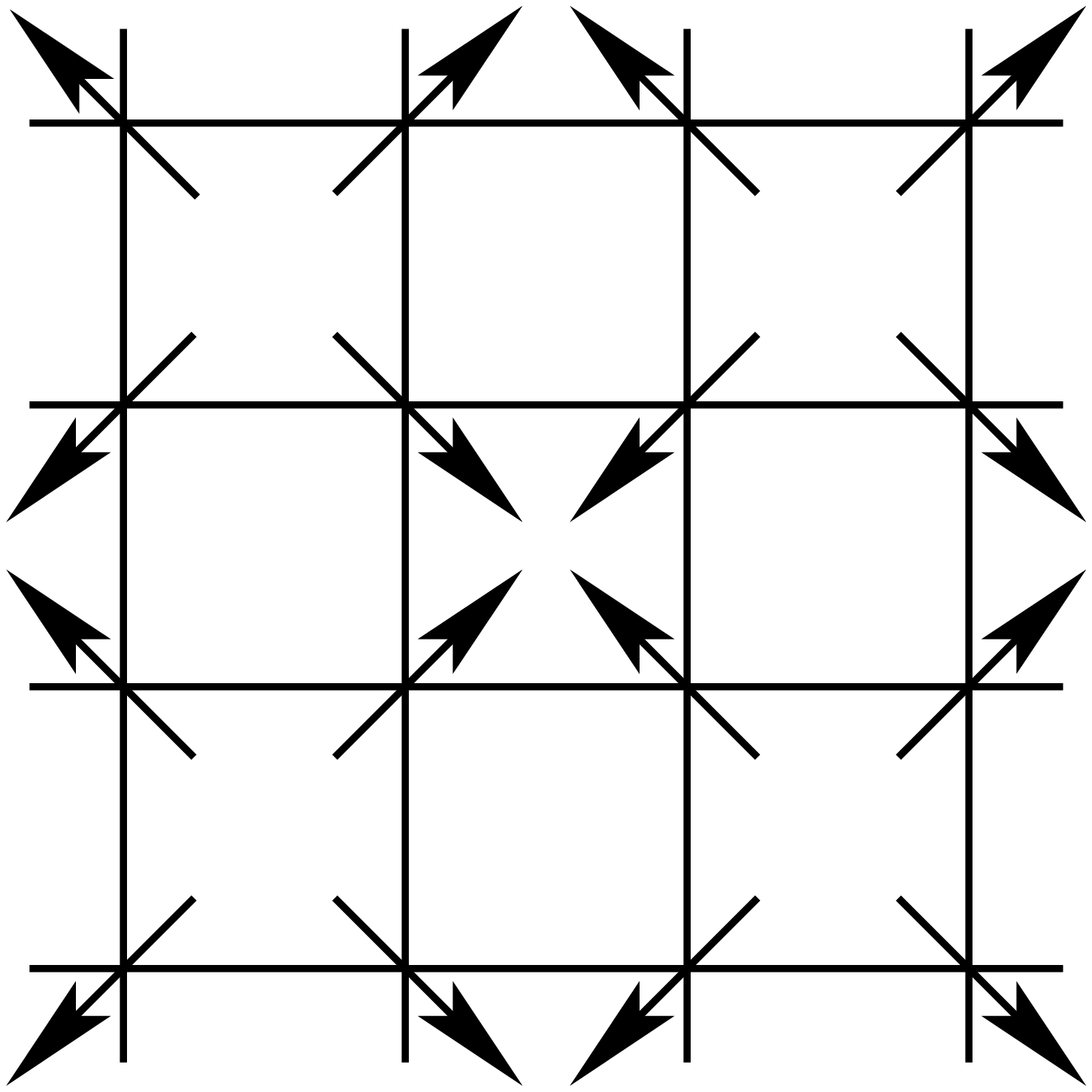}}
\mbox{\hspace{0.3cm}\includegraphics[width=3.4cm]{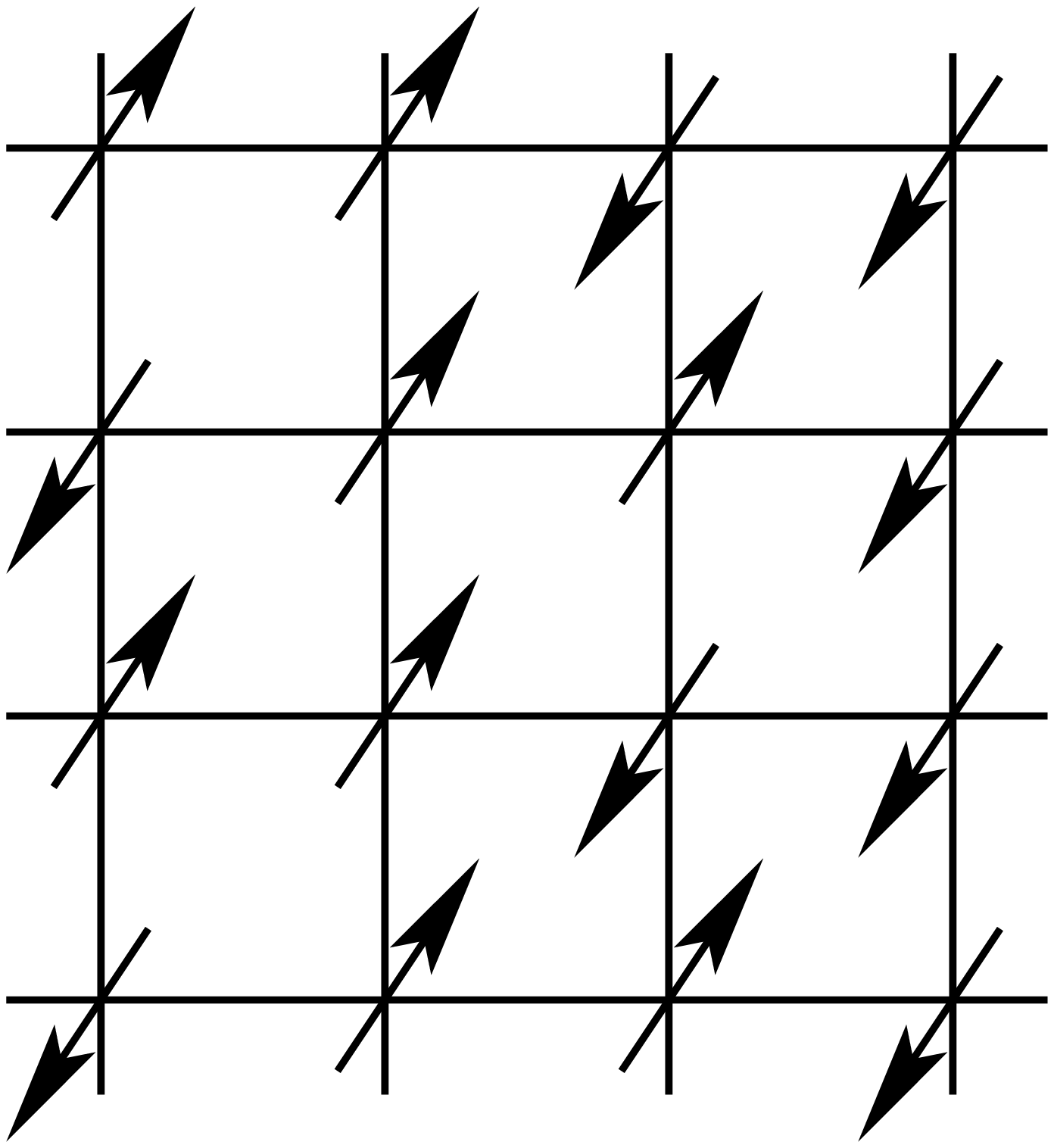}}
\centerline{ \quad (a) \qquad\qquad\qquad\qquad\qquad (b) }
\caption{(a) Schematic representation of a static spin configuration 
with vector-chirality order. (b) A ``three-up, one-down'' configuration, 
as obtained for $x = 0$ both in an unprojected model using $U = 10t$, 
$J = 4V = 0.01t$, and $K = 25t$, and in a fully projected model with 
$J = 4V = 0.01t$ and $K = 10t$.}
\label{flks}
\end{figure}

We thus conclude this section by commenting briefly on the effects of a 
very strong ring-exchange interaction. In spin systems, as realized at 
half-filling for large $U$, the ring-exchange term is generally considered 
to favor a maximization of the solid angle of the four spins on a plaquette, 
and thus to yield non-collinear magnetic phases. In the large-$K$ limit one 
expects the (static\cite{rcgb} or dynamic\cite{rgnb,rlst}) vector-chirality 
phase depicted in Fig.~\ref{flks}(a). This type of spin configuration has a 
low bond kinetic energy but a finite spin flux in the sense familiar from 
double-exchange models,\cite{rayanhaa} that a propagating electron picks 
up a path-dependent phase. In practice the bond current and kinetic 
energy vanish at half-filling in this static limit for large-$U$ systems. 
In the presence of holes a strong ring-exchange interaction is also 
expected (from the magnitude of the expectation values involved) to 
promote magnetic rather than bond order. 

However, in our analysis of the large-$K$ limit for an unprojected model 
with $U/t \gg 1$, both at and away from half-filling, we have found a 
different ground state, shown in Fig.~\ref{flks}(b).
This spin configuration ensures a ``three-up, one-down'' state on every 
plaquette, which gives the same ring-exchange energy per plaquette [see 
Eq.~(\ref{ehk})] as the vector-chirality configuration. It is not unique, 
and need not conserve a total spin component $S_z^{\rm tot} = 0$. The 
extra stability of the ``three-up, one-down'' phase may be ascribed 
primarily to the fact that, in approaching the limit of an effective 
spin model, residual dynamical processes act to favor collinear spin 
states. However, quantum fluctuations not contained at the Hartree-Fock 
level could also play an important role in the stability of the phases in 
Fig.~\ref{flks}.
 
In the projected model obtained by consistent derivation from the large-$U$ 
Hubbard Hamiltonian, we have been unable to find these states because 
of continued competition from superexchange terms, which are always 
significant. In this case the mutual frustration of the two competing 
interactions raises the possibility of further exotic spin configurations, 
such as the scalar-chirality phase,\cite{rlst,rgnb} or different dimerized 
structures breaking lattice translation symmetry.\cite{rckk,rsdss,rlst,rgnb}
However, with an arbitrary choice of large $K$ unrelated to the consistency 
considerations of Sec.~IV, the same ``three-up, one-down'' phase 
emerges directly [Fig.~\ref{flks}(b)].

\section{Summary and Conclusions}

In this contribution we have considered a microscopic description of 
circulating charge- and spin-current states in a planar model appropriate 
for cuprate systems, and the qualitative influence upon these phases
of cyclic, four-spin interactions. The CF and SF phases were established 
on a small cluster by iterative solution for the electronic states within 
the HFA for the generalized Hubbard model. Although the real-space 
Hartree-Fock approach does not in general favor homogeneous and dynamically 
stabilized states, prefering localized and charge-inhomogeneous structures, 
meaningful qualitative comparisons remain possible.

The robust qualitative conclusions concerning flux phases are the 
following. CF and SF states are favored for small on-site repulsion $U$, 
and are stabilized primarily by next-neighbor superexchange ($J$) and 
Coulomb ($V$) contributions. They are most favorable close to 
half-filling, becoming unstable to non-ordered metallic phases at  
higher hole doping ($x > 0.15$ in HFA), a result which reflects 
qualitatively the onset of the Fermi-liquid phase beyond optimal doping 
in the cuprates. The bond current $I$ increases with interaction 
strength, tending in the limit $|J|/t, 4V/t \gg 1$ towards a $\pi$-flux 
phase in the CF state, and in the SF state to an imaginary bond order 
parameter preceding a ferromagnetic instability. Very few inhomogeneous 
current states are found in this type of analysis, and coexisting states 
are dominated by localized charge and magnetic configurations. 

The inclusion of strong electronic correlations by projection of the 
hopping terms may yield solutions qualitatively different from those 
obtained for independent electrons in a weak-coupling treatment 
(unprojected hopping). Projected hopping is therefore essential 
in making contact with cuprate-relevant parameters. Inclusion of 
projection effects within the Gutzwiller approximation shows that the 
CF phase does become competitive for realistic cuprate interactions. 
The CF phase in the projected model has a clearly defined regime of 
stability at low and intermediate doping, in close correspondence to 
the underdoped and optimally doped regimes in cuprates. While projection 
produces very small kinetic terms near half-filling, it also pushes the 
general CF state towards a $\pi$-flux phase. Next-neighbor hopping 
contributions ($t'$) of the magnitude and sign relevant for cuprate 
systems appear to reinforce the projected CF phase. By contrast, the 
SF phase is excluded in a projected model, demonstrating that this 
type of flux phase is not compatible with strong electron correlations. 

Introduction of a ring-exchange interaction ($K > 0$) is detrimental to 
the stability of CF phases in both unprojected and projected models. By
contrast, the SF phase shows an enhanced current and minimal energy 
penalty in the unprojected framework. Ring-exchange effects vary little 
with doping until the spin and bond order parameters are strongly 
reduced at high filling. The parameter values in cuprates are such 
that the $K$ term is largely irrelevant for qualitative phenomena at 
the Hartree-Fock level, and its effect remains weak even with the strong 
enhancement provided by the Gutzwiller approximation. However, we have 
shown that an exception to this statement arises for ordered magnetic 
states at very low filling, where the ring-exchange interaction might 
become qualitatively significant in a projected model. 

\acknowledgments

We thank C. Balseiro, A. L\"auchli, M. Randeria, K. Ro\'sciszewski, 
and M. Vojta for valuable discussions. This work was supported by the 
Swiss National Science Foundation and by the Polish State Committee 
of Scientific Research (KBN) under Project No.~1 P03B 068 26.

\end{document}